\documentclass[preprint,floats,nofootinbib,superscriptaddress,color]{revtex4}
\usepackage{dcolumn}   % needed for some tables
\usepackage{bm}        % for math
\usepackage{multirow}

\usepackage{amsmath}
\usepackage{amsfonts}
\usepackage{amssymb}
\usepackage{makeidx}
\usepackage{graphicx}
\usepackage{anysize}

\newcommand{\bs}{\boldsymbol}

\newcommand{\ba}{\begin{eqnarray}}
\newcommand{\ea}{\end{eqnarray}}
\newcommand{\be}{\begin{equation}}
\newcommand{\ee}{\end{equation}}

\newcommand{\pnslash}{\not{\!P_N}}
\newcommand{\pbslash}{\not{\!\overline{P}}}

\newcommand{\slslash}{\not{\!S_L}}

\usepackage[dvips]{color}

\begin{document}
\title{Parity violation and dynamical relativistic effects in $(\vec{e},e'N)$ reactions}

\author{R.~Gonz\'alez-Jim\'enez}
\affiliation{Departamento de F\'{\i}sica At\'{o}mica, Molecular y Nuclear, Universidad de Sevilla, 41080 Sevilla, Spain}
\affiliation{Department of Physics and Astronomy, Ghent University, Proeftuinstraat 86, B-9000 Gent, Belgium}

\author{J.A.~Caballero}
\affiliation{Departamento de F\'{\i}sica At\'{o}mica, Molecular y Nuclear, Universidad de Sevilla, 41080 Sevilla, Spain}

\author{T.W.~Donnelly}
\affiliation{Center for Theoretical Physics, Laboratory for Nuclear Science and Department of Physics,
Massachusetts Institute of Technology, Cambridge, Massachusetts 02139, USA}

\date{\today}

\begin{abstract}

It is well known that coincidence quasielastic $(\vec{e},e'N)$ reactions are not appropriate to analyze effects linked to 
parity violation due the presence of the fifth electromagnetic (EM) response $R^{TL'}$. Nevertheless, in this work we develop
a fully relativistic approach to be applied to parity-violating (PV) quasielastic $(\vec{e},e'N)$ processes. This is of importance
as a preliminary step in the subsequent study of inclusive quasielastic PV $(\vec{e},e')$ reactions. Moreover, our present
analysis allows us to disentangle effects associated with the off-shell character of nucleons in nuclei, gauge ambiguities and
the role played by the lower components in the nucleon wave functions, {\it i.e.,} dynamical relativistic effects. This study can help
in getting clear information on PV effects. Particular attention is paid to the relativistic plane-wave impulse approximation where
the explicit expressions for the PV single-nucleon responses are shown for the first time.

\end{abstract}

% \pacs{12.15.-y, 12.15.Lk, 12.15.Mm, 14.20.Dh, 14.65.Bt, 25.30.Bf}
% % PACS numbers:
%
% % 12.15.-y  Electroweak interactions
% % 12.15.Lk Electroweak radiative corrections
% % 12.15.Mm Neutral currents
% % 14.20.Dh Protons and neutrons
% % 14.65.Bt Light quarks
% % 25.30.Bf Elastic electron scattering
\maketitle

\tableofcontents

\section{Introduction}

The electroweak structure of the nucleon can be parameterized in terms of nine form factors, 
three for each flavor ($u$, $d$, $s$) corresponding to the electric (E), magnetic (M) and axial-vector (A) sectors. 
The sole use of PV electron-proton (PVep) asymmetry measurements does not allow one to extract the 
nine form factors of the nucleon. On the contrary, the combination of parity-conserving (PC) cross sections and 
parity-violating (PV) asymmetries in elastic and quasielastic (QE) electron scattering processes, in addition 
to measurements of different observables from neutrino scattering and beta decay, constitutes the general framework in which the determination 
of the form factors of the nucleon should be accomplished.

PV electron scattering is a powerful tool to study the weak neutral current (WNC) and it can provide useful information on the strange
matrix elements ($\bar{s}\gamma_\mu s$ and $\bar{s}\gamma_\mu\gamma_5 s$) in nucleons and nuclei. 
Strange form factors contain new information (additional to the electromagnetic (EM) ones) on the nucleon structure, 
and provide also strong constraints to any 
microscopic model aiming to describe the nucleonic structure starting from Quantum Chromodynamics (QCD).

It has been proven~\cite{Gonzalez-Jimenez13a} that the PVep asymmetry is an excellent observable in order to determine 
the vector strange form factors of the nucleon. 
However, this requires one to have good knowledge of the remaining ingredients that enter in the description of the asymmetry, in particular, 
the EM and axial-vector form factors and the WNC effective weak coupling constants (that include radiative corrections).
With regards to the EM form factors, their general structure and behavior are well described. 
In the case of the axial-vector form factor, most of the information we have comes from the analysis of neutrino scattering experiments 
and beta-decay measurements. 
Although some discussion has recently emerged on the value of the axial-vector mass due to the data taken by the MiniBooNE 
collaboration~\cite{MiniBooNECC10,MiniBooNENC10} (see also~\cite{Gonzalez-Jimenez13b,Megias13} and references therein), 
the {\it standard} parameterization of the axial-vector form factor is still accepted by the majority of the scientific community.
However, a serious problem which is not yet solved concerns the treatment of radiative corrections. 
Authors in~\cite{Musolf90,Horowitz93a} claim 
that radiative corrections are very small for processes where only the weak coupling takes place. 
Hence, the description of the axial-vector form factor at tree-level is expected to be a very good approximation in reactions 
that involve neutrinos/antineutrinos as probes. 
On the contrary, radiative corrections in the axial-vector sector for PV electron scattering are very important (in contrast 
to the purely vector current). 
This is one of the main sources of error for the determination of the strange magnetic form factor through the analysis of PVep asymmetry data. 
The strong correlation between the magnetic and electric strangeness content in the nucleon ($\mu_s$ and $\rho_s$) 
leads the previous uncertainties to be propagated also to the strange electric form factor.
The main contribution in the axial-vector form factor comes from the isovector ($T=1$) channel; therefore, the evaluation and knowledge 
of the isovector contribution to the axial nuclear response, $R_A^{T=1}$, is of great importance in order to interpret correctly the PV asymmetry. 
Nowadays it constitutes one of the main challenges both experimentally and theoretically to the scientific community. 

This work deals with the study of exclusive PV electron-nucleus scattering processes. 
We restrict ourselves to the QE regime that corresponds to the electron being scattered from a single nucleon that is subsequently 
ejected from the target nucleus and detected in coincidence with the scattered electron. 
The analysis of exclusive $({\vec e},e'N)$ processes
constitutes a preliminary step in the study of PV effects in inclusive $({\vec e},e')$ reactions. 
The latter are of great relevance
in order to get more insight into the weak structure of the nucleon. In $(\vec{e},e'N)$ reactions the
description of final-state interactions (FSI) between the ejected nucleon and the residual nucleus 
leads to the appearance of the so-called fifth EM response function $R^{TL'}$.
This response only contributes if the helicity of the incident electron is measured and, consequently, its contribution to
the PV asymmetry is different from zero.
This result is of great importance because it might lead the exclusive parity-violating quasielastic (PVQE) 
asymmetry to be irrelevant when attempting to get information on the 
responses attached to the interference between EM and WNC currents (henceforth simply denoted as {\it PV responses}). 
Note that the EM contributions are in general several orders of magnitude larger than the interference ones.
However, there are some particular kinematics for which the contribution of the PV responses 
can be similar or even larger than the purely EM one.
Moreover, this study provides useful information on the uncertainties linked to the treatment of the off-shell vertex and on the discrepancies 
associated with the use of different nuclear models. 
These effects can be of importance in the subsequent analysis of inclusive $(\vec{e},e')$ 
processes where the fifth EM response does not enter. Hence, the work presented here should be considered in concert with the study
applied to inclusive QE electron scattering that is presented in an accompanying paper~\cite{inclusive}. 
Whereas here we emphasize the role
played by dynamical relativistic effects, gauge ambiguities and off-shell uncertainties 
in the {\it exclusive} PV responses,
being aware of the enormous difficulties in getting information on PV effects, in \cite{inclusive} the focus is placed on the behavior
of the {\it inclusive} PV responses and the asymmetry, and how they are affected by the weak structure of the nucleon.

To conclude, we summarize in what follows how this paper is organized.
In Sect.~\ref{SEDchapter} we introduce the general formalism needed to evaluate the exclusive cross section for PVQE electron-nucleus scattering.
We start by describing the kinematics and the calculation of the differential cross section for the exclusive process. 
Then we present the models and approximations 
employed to the description of the nuclear vertex. 
Special emphasis is placed on the general formalism involved 
in the relativistic plane-wave impulse approximation (see Sect.~\ref{RPWIAsect}).
Section~\ref{resultados-excl} presents the analysis of the results.
The effects in the PV responses due to the use of different prescriptions for the nuclear current and the treatment 
of FSI are analyzed in Sects.~\ref{exclResp} and \ref{excluFSI}, respectively.
In Sect.~\ref{AsiExcl} we study the helicity asymmetry linked to the exclusive process.
There we study the impact on the asymmetry due to FSI, relativistic dynamical effects, off-shell effects
and the particular description of the form factors of the nucleon.
Finally, a brief summary and our main conclusions are presented in Sect.~\ref{conclusionsIII}.

\section{Formalism for $(\vec{e},e'N)$ reactions with Parity Violation}
\label{SEDchapter}

\begin{figure}[htbp]
    \centering
        \includegraphics[width=.6\textwidth,angle=0]{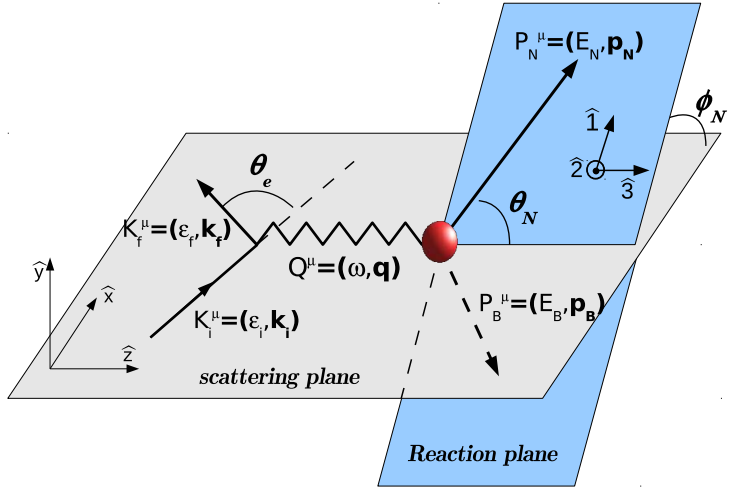}
    \caption{(Color online) General scheme of the scattering process $A(e,e'N)B$. The scattering frame is defined
      by $\lbrace \hat{{\bs x}},\hat{{\bs y}},\hat{{\bs z}}\rbrace$, whereas the hadronic reference system is
      given by $\lbrace \hat{{\bs 1}},\hat{{\bs 2}},\hat{{\bs 3}}\rbrace$. Also shown are the four-momenta and angular variables
      that enter in the description of the process.}
    \label{fig:planos2}
\end{figure}

In this section we summarize the general formalism involved in the description of $(\vec{e},e'N)$ reactions
when the weak interaction is included in addition to the dominant EM one.
We use the Born approximation, that is, 
only one boson, photon for the EM process and $Z$-boson for the weak interaction, is exchanged. A general
representation of the process is illustrated in Fig.~\ref{fig:planos2}. Here the incident electron,
with four-momentum $K_i^\mu=(\varepsilon_i,{\bf k}_i)$ and helicity $h$, is scattered through an angle $\theta_e$ to
four-momentum $K_f^\mu=(\varepsilon_f,{\bf k}_f)$. The nuclear target is characterized in the lab frame by
$P_A^\mu=(M_A,{\bf 0})$ and the residual system by $P_B^\mu=(E_B,{\bf p}_B)$. The four-momentum corresponding to the
ejected nucleon is denoted as $P_N^\mu=(E_N,{\bf p}_N)$. Finally, the transferred four-momentum in the process (carried
by the photon or $Z$-boson) is given by $Q^\mu=(\omega,{\bf q})$.

Detailed studies of the general kinematics involved in exclusive $(e,e'N)$ reactions have been presented in previous works~\cite{Caballero98a,Caballero98b}.
Here we simply recall the basic quantities of interest for later discussion. The missing momentum ${\bf p}_m$ is defined as
${\bf p}_m\equiv -{\bf p}_B={\bf p}_N-{\bf q}$. The magnitude $p_m=|{\bf p}_m|$ characterizes the split in momentum flow between the 
detected nucleon and the unobserved daughter nucleus. Correspondingly, an excitation energy of the residual system can be 
introduced: $\varepsilon\equiv E_B-E_B^0$ with $E_B^0$ the total energy of the residual nucleus in its ground state.

The $(\vec{e},e'N)$  process is completely determined by six independent kinematical variables. 
From these, the dependence on the electron scattering angle $\theta_e$
and the azimuthal angle $\phi_N$ can be isolated by geometry. 
In contrast, the dependences on the four remaining variables (denoted as dynamical
variables) involve detailed aspects of the nuclear current matrix elements. Notwithstanding, energy and momentum conservation can be used to
determine the allowed regions in the $(\varepsilon - p_m)$ plane (the interested reader can go to \cite{Caballero10b,Donnelly99b,Martinez02b} for details). 
Once the general kinematics have been set up, the general cross section in the laboratory system can be written as~\cite{Raskin88,Caballero98a,Martinez02b},
\begin{eqnarray}
 \dfrac{d\sigma}{d\varepsilon_f d\Omega_f  d\Omega_N } 
 = \frac{K}{8\pi^2} f_{rec}^{-1}\left(\frac{\varepsilon_f}{\varepsilon_i}\right)
\left[\frac{1}{Q^4}\eta_{\mu\nu}W^{\mu\nu} 
 + \frac{-2}{Q^2M_Z^2}{\cal R}e
  \left(\widetilde{\eta}_{\mu\nu}\widetilde{W}^{\mu\nu}\right)\right] \, ,
\end{eqnarray}
where we have introduced the kinematical constant $K\equiv\frac{M_BM_Np_N}{M_A}$ and $f_{rec}$ is the usual recoil factor~\cite{Martinez02b}.
The purely EM and interference leptonic tensors are given by
\begin{eqnarray}
\eta_{\mu\nu}&=&e^2\left(s_{\mu\nu}+ha_{\mu\nu}\right) \\
\widetilde{\eta}_{\mu\nu}&=&\frac{eg}{4\cos\theta_W}\left[
(a_V-ha_A)s_{\mu\nu}+(ha_V-a_A)a_{\mu\nu}\right] 
\end{eqnarray}
with $a_V$ and $a_A$ the vector and axial-vector WNC electron couplings, 
$h$ the electron helicity and we have separated the overall tensor into
its symmetric ($s_{\mu\nu}$) and antisymmetric ($a_{\mu\nu}$) contributions:
\be
s_{\mu\nu}= K^i_\mu K^f_\nu+K^i_\nu K^f_\mu+\frac{Q^2}{2}g_{\mu\nu} \, , \,\,\,\,\,\,\,
a_{\mu\nu}= i\epsilon_{\mu\nu\alpha\beta}K_i^\alpha K_f^\beta \, .
\ee

The hadronic tensors contain all of the information on the nuclear structure, and they are given as
\begin{eqnarray}
W^{\mu\nu}
     =\overline{\sum_{IF}}(J_{EM}^\mu({\bf q}))^*J_{EM}^\nu({\bf q})\,,
     \,\,\,\,\,\,\,\,\,\,\,\,\,\,\,
%%%%%%%%%%%%%%%
\widetilde{W}^{\mu\nu} 
   = \overline{\sum_{IF}}(J_{EM}^{\mu}({\bf q}))^*J^{\nu}_{WNC}({\bf q})\,,
   \label{WgWz} 
\end{eqnarray}
in terms of the purely EM and WNC matrix elements:
\begin{eqnarray}
J_{EM}^\mu({\bf q}) = e \langle N;B|\hat{J}_{EM}^\mu| A\rangle\,,
\,\,\,\,\,\,\,\,\,\,\,\,\,\,\,
J_{WNC}^\mu({\bf q}) = \left(\frac{eg}{\cos\theta_W}\right) 
           \langle N;B| \hat{J}_{WNC}^\mu| A\rangle\, .
\end{eqnarray}
The contraction of the leptonic and hadronic tensors can be expressed in terms of six response functions that are given 
by taking the appropriate components of the hadronic tensors (see~\cite{Donnelly85} for details):
\begin{equation}
 \eta_{\mu\nu}W^{\mu\nu} 
          =  e^2 2v_0\biggl[v_LR^L + v_T R^T + v_{TL}R^{TL} + v_{TT}R^{TT}
          +hv_{TL'}R^{TL'}\biggr]\,, \label{lWEM}
\end{equation}
where $v_0=4\varepsilon_i\varepsilon_f\cos^2\frac{\theta_e}{2}$ and the labels $L$ and $T$ refer to projections of 
the current matrix elements longitudinal and transverse to the direction of the momentum carried by the exchanged virtual boson. 
The terms $v_\alpha$ denote the kinematical factors that depend only on the leptonic
tensors and whose explicit expressions are given in \cite{Donnelly85}. 

Likewise, the contraction of the $\gamma-Z$ interference tensors can be written in the form
\begin{eqnarray}
 \widetilde{\eta}_{\mu\nu}\widetilde{W}^{\mu\nu} 
%%%%%%%%%%%%%%%%%              
          &=&\frac{-1}{8m^2}\left(\frac{eg}{\cos\theta_W}\right)^2
          2v_0\left[(a_V-ha_A)\left(v_L\widetilde{R}^L + v_T \widetilde{R}^T 
               + v_{TL}\widetilde{R}^{TL} + v_{TT}\widetilde{R}^{TT}\right)\right.\nonumber\\ 
          &+& \left.(ha_V-a_A) \left(v_{T'}\widetilde{R}^{T'} +
          v_{TL'}\widetilde{R}^{TL'}\right)\right]\,.       \label{lWWNC}  
\end{eqnarray}

The general expression for the exclusive cross section in presence of the weak interaction finally results:
\begin{eqnarray}
& & \dfrac{d\sigma}{d\varepsilon_f d\Omega_f  d\Omega_N } 
  = \sigma_{M}Kf_{rec}^{-1}\Biggl\lbrace  
   \sum_{\alpha=L,T,TL,TT} v_\alpha R^\alpha + hv_{TL'}R^{TL'} \nonumber \\
&&  -\frac{{\cal A}_0}{2}\biggl[
     (a_V-ha_A)\sum_{\alpha=L,T,TL,TT} v_\alpha \widetilde{R}^\alpha
  + (ha_V-a_A)\sum_{\alpha'=T',TL'} v_{\alpha'}\widetilde{R}^{\alpha'}
  \biggr]\Biggr\rbrace\, ,
    \label{sigexclR1}
\end{eqnarray}
where we have introduced the Mott cross section $\sigma_{M}=\dfrac{4\alpha^2}{Q^4}\varepsilon_f^2\cos^2(\theta_e/2)$, and
the term ${\cal A}_0$ that scales the PV effects:
\begin{eqnarray}
-\dfrac{{\cal A}_0}{2} \equiv 
\dfrac{2 Q^2}{e^2M_Z^2}\left(\dfrac{g}{4\cos\theta_W}\right)^2\,.\label{A0}
\end{eqnarray}

The evaluation of the EM and PV hadronic response functions ($R^\alpha, \widetilde{R}^\alpha$) requires the knowledge
of the corresponding nuclear tensors: $W^{\mu\nu}$, $\widetilde{W}^{\mu\nu}$. This implies a description of the nuclear
initial and final states and the many-body current operators. This is a very complicated, almost unapproachable, problem
unless specific approximations are considered. In our case, we focus on the kinematical region close to the QE
peak where the impulse approximation (IA) constitutes an excellent description of the problem. Within the IA the
exchanged boson (photon and/or $Z$) interacts only with one nucleon that is consequently ejected. Hence the scattering process
is given simply as an incoherent sum of single-nucleon scattering processes, {\it i.e.,} the current is taken as a one-body operator 
and one makes use of single-nucleon wave functions. In momentum space we may write in general
\be
J^\mu\equiv\int d{\bf p}\ \overline{\Phi}_F({\bf p}+{\bs q})
    \hat{J}^\mu\Phi_B({\bf p})\, ,\label{JFIp}
\ee
where $\Phi_B$ and $\Phi_F$ are the bound and scattered nucleon wave functions, respectively, and 
$\hat{J}^\mu_N$ the one-body nucleon current operator. 

Within the IA the virtual boson 
attaches to a single bound nucleon with four-momentum $P^\mu=(E,{\bf p})$ that is consequently ejected
from the nucleus and interacts with the residual nucleus. The asymptotic four-momentum of the nucleon is
given by $P_N^\mu=(E_N,{\bf p}_N)$, and the residual nucleus is characterized by $P_B^\mu=(E_B,{\bf p}_B)$.

In this work we use a fully relativistic calculation where the bound nucleon states are given as self-consistent Dirac-Hartree solutions, 
derived within a relativistic mean field (RMF) approach using a Lagrangian containing $\sigma$, $\omega$, and $\rho$ mesons~\cite{Sharma93}. 
The ejected nucleon state is described as a relativistic scattering wave function. 
Here different options have been considered. First, the relativistic 
plane-wave impulse approximation (RPWIA),
namely, the use of relativistic plane-wave spinors, {\it i.e.,} no interaction between the outgoing nucleon and the residual
nucleus is considered; second, the effects of FSI are incorporated by solving the Dirac equation
in the presence of relativistic optical energy-dependent scalar and vector potentials. This constitutes the relativistic distorted-wave impulse approximation (RDWIA). In this work we make use of the particular prescription EDAI-O for the optical potential
(see \cite{Schwandt82,Comfort80,Hama90,Cooper93} for details). 

Concerning the current operator, we use the relativistic free nucleon expressions for the two usual prescriptions considered
in the literature, {\it i.e.,} $CC1$ and $CC2$~\cite{Forest83}:

%%%%%%%%%%%%%%%%%%%%%%%%%%%%%%%%%%%%%%%%%%%%%%%%%%%%%%%%%%%%%%%%%%%%%%%%%%%%%
%%%%%%%%%%%%%%%%%%%%%%%%%%%%%%%%%%%%%%%%%%%%%%%%%%%%%%%%%%%%%%%%
\begin{itemize}
 \item {\sc Electromagnetic Current Operator:}
\begin{eqnarray}
\left. \hat{J}^{\mu}_{EM}\right|_{CC1}  &=& 
(F_1+F_2)\gamma^{\mu}-\frac{F_2}{2M_N}(\overline{P}+P_N)^\mu \, ,\label{CC1em} \\
\left. \hat{J}^{\mu}_{EM}\right|_{CC2} &=& 
F_1\gamma^{\mu}+i\frac{F_2}{2M_N}\sigma^{\mu\nu}Q_{\nu}\, .\label{CC2em}
\end{eqnarray}
\item {\sc Vector Neutral Current:}
\begin{eqnarray}
\left. \hat{J}^{\mu}_{WNC,V}\right|_{CC1}  &=&
(\widetilde{F}_1+\widetilde{F}_2)\gamma^{\mu}-\frac{\widetilde{F}_2}{2M_N}(\overline{P}+P_N)^{\mu}\,,
\label{CC1dv}\\
\left. \hat{J}^{\mu}_{WNC,V}\right|_{CC2} &=& 
\widetilde{F}_1\gamma^{\mu}+i\frac{\widetilde{F}_2}{2M_N}\sigma^{\mu\nu}Q_{\nu}\, .
\label{CC2dv}
\end{eqnarray}
\item {\sc Axial Neutral Current:}
\begin{eqnarray}
\hat{J}^{\mu}_{WNC,A} = G_A^e\gamma^{\mu}\gamma^5 + \frac{\widetilde{G}_P}{M_N}Q^{\mu}\gamma^{5} \, .\label{opAxial}
\end{eqnarray}
\end{itemize}
We have introduced the {\it ``on-shell''} four-momentum $\overline{P}^\mu=(\overline{E},{\bf p})$ 
with $\overline{E} =\sqrt{p^2 + M_N^2}$ and ${\bf p}$ the bound nucleon momentum. Note that the two
prescriptions, $CC1$ and $CC2$ are equivalent for free on-shell nucleons. However, the IA deals with off-shell
bound and ejected nucleons. Hence the two operators lead to different results. Moreover, current conservation
(EM and vector WNC contributions) is in general not fulfilled and consequently uncertainties 
dealing with the particular gauge selected also emerge. Here we consider three different options:
i) no current conservation is imposed at all (Landau gauge), ii) current conservation is imposed
by eliminating the third component (Coulomb gauge), and iii) as in the previous case but eliminating the time
component (Weyl gauge). In next sections we estimate and analyze the uncertainties introduced by these different
options in the PV responses and in the PVQE asymmetry.

\subsection{Relativistic Plane-Wave Impulse Approximation (RPWIA)}
\label{RPWIAsect}

In this section we present the response functions and cross section within the RPWIA, namely, neglecting FSI 
between the outgoing nucleon and the residual nucleus. Although this is an oversimplified description of the scattering process, it
allows one to get analytical expressions for the responses, hence providing significant insight into the specific behavior of
the various observables. A great advantage of the RPWIA is linked to the clear separation between the contributions associated with
the upper and lower components in the relativistic bound nucleon wave functions. This is known as {\sl ``relativistic dynamics''} 
or {\sl ``spinor distortion''} in contrast to the purely relativistic kinematical effects. In what follows we briefly present the general procedure of the analysis. 
We follow closely our previous studies in \cite{Caballero98a,Martinez02a,Martinez02b} and give all details in Appendix~\ref{apendiceRPWIA}.

The hadronic current in RPWIA is given in the general form:
\be
J^\mu = \overline{u}_N({\bf p}_N,s_N)\hat{J}^\mu\Phi_\kappa^m({\bf p}) \,,
 \label{JFIpw} 
\ee
where $u({\bf p}_N,s_N)$ is a Dirac free spinor describing the outgoing nucleon, whereas 
$\Phi_\kappa^m({\bf p})$ is the Fourier transform of the bound nucleon relativistic wave function evaluated with the RMF model. 
This 4-component wave function can be expressed in terms of the
free Dirac spinors: $u({\bf p},1/2)$, $u({\bf p},-1/2)$, $v({\bf p},1/2)$ and $v({\bf p},-1/2)$.
Proceeding this way, one can identify the specific contributions associated with the upper (positive-energy)
and lower (negative-energy) components in the response functions and the cross section. 
The general (non-trivial) procedure has been presented in \cite{Caballero98a} in the case of PC electron scattering processes, namely, for EM response functions. 
In this section we extend the analysis to the PV responses and obtain the final expressions that will be of interest for the discussion of results in the
next section. 
As already mentioned, all details on the developments concerning the nucleonic tensors are given in Appendix~\ref{apendiceRPWIA}.

The hadronic EM and interference WNC matrix elements in RPWIA result
\begin{eqnarray}
  J^{\mu}_{EM}&=&e[\overline{u}({\bf p}_N,S_N)\hat{J}^{\mu}_{EM}\ 
  \Phi_k^{m_j}({\bf p})]\, ,\label{JEMh}\\
  J^{\mu}_{WNC}&=&\left(\frac{g}{4\cos\theta_W}\right)[\overline{u}({\bf p}_N,S_N)
\hat{J}^{\mu}_{WNC}\ \Phi_k^{m_j}({\bf p})]\label{JZh}\, . 
\end{eqnarray}

After laborious algebra (see Appendix~\ref{apendiceRPWIA}), the hadronic tensors can finally be written in the form
\begin{eqnarray}
 W^{\mu\nu} &=&
 e^2 \biggl({\cal W}^{\mu\nu}N_{uu}(p) 
 + {\cal Z}^{\mu\nu}N_{vv}(p) 
 + {\cal N}^{\mu\nu}N_{uv}(p)\biggr)\, , \\
 %%%%%%%%%%%%%%
\widetilde{W}^{\mu\nu} &=& \left(\frac{eg}{4\cos\theta_W}\right)
\biggl(\widetilde{{\cal W}}^{\mu\nu}N_{uu}(p) 
+ \widetilde{{\cal Z}}^{\mu\nu} N_{vv}(p)
+\widetilde{{\cal N}}^{\mu\nu} N_{uv}(p)\biggr)\, ,
\end{eqnarray}
where in both tensors we have isolated the contribution coming from the positive-energy components (denoted by the indices $uu$),
the negative-energy ($vv$ term) and the interference ones ($uv$). Note that the three terms factorize into single-nucleon tensors multiplied by 
functions associated with the upper ($u$) and lower ($v$) components in the bound nucleon wave function in momentum space. 
These functions can be interpreted as the positive-energy ($N_{uu}$), negative-energy ($N_{vv}$) and interference ($N_{uv}$) contributions to the nucleon momentum distribution. 
The explicit expressions are given in Appendix~\ref{apendiceRPWIA} (see also \cite{Caballero98a,Caballero98b} for more details).

With regards to the single-nucleon tensors, the ones corresponding to the purely EM sector, ${\cal W}^{\mu\nu}$, ${\cal Z}^{\mu\nu}$ and ${\cal N}^{\mu\nu}$,
have been analyzed in detail in~\cite{Caballero98a} providing explicit expressions for the two current operator prescriptions considered. 
The Lorentz invariant EM amplitude is given by\footnote{Note that the fifth EM response, $R^{TL'}$, does not enter in RPWIA.}
\be
 \eta_{\mu\nu}W^{\mu\nu} \propto 
    s_{\mu\nu}\left[ {\cal W}^{\mu\nu}N_{uu}(p) 
    + {\cal Z}^{\mu\nu}N_{vv}(p) + {\cal N}^{\mu\nu}N_{uv}(p)\right]\, .
\ee

By contrast with the EM case~\cite{Caballero98a}, the single-nucleon electroweak interference tensors,
$\widetilde{{\cal W}}^{\mu\nu}$, $\widetilde{{\cal Z}}^{\mu\nu}$ and $\widetilde{{\cal N}}^{\mu\nu}$, present a rather more complex structure with its symmetrical and antisymmetrical contributions not so clearly isolated
(see Appendix~\ref{apendiceRPWIA} for explicit expressions). 
However, since the purely vector component in the WNC (denoted by the index V)
leads to purely real tensors, whereas the axial term (denoted by A) gives purely imaginary tensors, 
it can be easily proven that its contraction with the corresponding leptonic tensor can be written as
\begin{eqnarray}
 {\cal R}e \left[\widetilde{\eta}_{\mu\nu}\widetilde{W}^{\mu\nu}\right] 
 &\propto& (a_V-ha_A)s_{\mu\nu}\left(
   \widetilde{{\cal W}}_V^{\mu\nu}N_{uu}(p) 
 + \widetilde{{\cal Z}}_V^{\mu\nu}N_{vv}(p) 
 + \widetilde{{\cal N}}_V^{\mu\nu}N_{uv}(p)\right)\nonumber\\
 %%%%%%%%%%%
&+& (ha_V-a_A)a_{\mu\nu}\left(
    \widetilde{{\cal W}}^{\mu\nu}_A N_{uu}(p) 
  + \widetilde{{\cal Z}}^{\mu\nu}_A N_{vv}(p) 
  + \widetilde{{\cal N}}^{\mu\nu}_A N_{uv}(p)\right)\, .
\end{eqnarray}
Note that only the symmetric contribution in the vector-type single-nucleon tensors and likewise, the antisymmetric axial-type tensor, contribute
to the cross section and response functions.

The hadronic responses in RPWIA in Eq.~(\ref{sigexclR1}) are built directly from the corresponding single-nucleon responses ${\cal R}^K_x$ multiplied by
the momentum distributions $N_{x}$ with $x=uu,uv,vv$. The differential cross section is finally given as
\begin{eqnarray}
&& \dfrac{d\sigma}{d\varepsilon_f d\Omega_f  d\Omega_N } 
    = \sigma_{M}\frac{M_BM_Np_N}{M_A\ f_{rec}} \sum_{x=uu,uv,vv} \Biggl\lbrace  
    \sum_{\alpha=L,T,TL,TT}v_\alpha{\cal R}^\alpha_{x}N_{x}(p)\nonumber\\
%%%%%%%    
      &-& \frac{{\cal A}_0}{2}\biggl[ (a_V-ha_A)
       \sum_{\alpha=L,T,TL,TT} v_\alpha\widetilde{{\cal R}}^\alpha_{x}N_{x}(p)
    + (ha_V-a_A)\sum_{\alpha'=T',TL'}v_{\alpha'}\widetilde{{\cal R}}^{\alpha'}_{x}
    N_{x}(p)\biggr]\Biggr\rbrace\,. \nonumber \\
    &&    \label{sigexclPW}
\end{eqnarray}

%%%%%%%%%%%%%%%%%%%%%%%%%%%%%%%%%%%%%%%%%%%%%%%%%%%%%%%%%%%
%%%%%%%%%%%%%%%%%%%%%%%%%%%%%%%%%%%%%%%%%%%%%%%%%%%%%%%%%%%

In the next section we analyze the results obtained for the interference response functions and PV asymmetry within RPWIA, 
taking the two prescriptions of the current operator and the three gauges. These results are also compared with
more sophisticated calculations where FSI have been incorporated through the use of relativistic complex optical potentials.
We discuss whether the introduction of the exclusive helicity asymmetry makes sense in getting information on PV effects,
and analyze their limits of applicability.

%%%%%%%%%%%%%%%%%%%%%%%%%%%%%%%%%%%%%%%%%%%%%%%%%%%%%%%%%%%
%%%%%%%%%%%%%%%%%%%%%%%%%%%%%%%%%%%%%%%%%%%%%%%%%%%%%%%%%%%

\section{Analysis of the results}
\label{resultados-excl}

In this section we present our results for the different exclusive observables: 
responses, cross section and helicity asymmetry. Our interest
is focused on the WNC interference contributions and how these can alter the purely EM responses. 
It is well known that the process
$(\vec{e},e'N)$  is not well suited to studying PV effects 
because it is very hard to devise an observable where the EM contributions, orders of 
magnitude larger than the PV interference ones, could be almost cancelled out. 
In fact, this is the reason to introduce the helicity asymmetry
in inclusive $(\vec{e},e')$  processes. 
However, whereas the difference between the $(\vec{e},e')$  cross sections 
corresponding to opposed electron helicities is non-zero because of the PV effects
(for the EM interaction with parity conservation such a difference is strictly zero), 
the situation is different for $(\vec{e},e'N)$  reactions. 
Here, the presence of FSI leads to the appearance of the so-called fifth response that is also 
linked to the electron helicity. This response is a purely
EM contribution to the helicity asymmetry and hence makes it very difficult to isolate contributions coming
from PV effects. 
However, we are convinced that the study of PV $(\vec{e},e'N)$  reactions 
might be of great interest as a preliminary
step in the subsequent study of inclusive processes. 
Moreover, off-shell and gauge ambiguities in $(\vec{e},e'N)$  also have an impact
on the inclusive responses and PV helicity that needs to be carefully evaluated. 
Therefore, in this paper we focus on the interference
$(\vec{e},e'N)$  observables paying special attention to the contribution 
of the positive- and negative-energy components in the nucleon
wave functions, and to the effects introduced by a proper description of the FSI between the outgoing nucleon and the
residual nucleus. The scattering reaction formalism is described fully relativistically, namely, not only are the kinematics relativistic, but
also the nuclear dynamics are described making use of the relativistic Dirac equation in the presence of relativistic potentials.

All results presented in the next sections correspond to $(q,\omega)$-constant kinematics (sometimes also referred as quasi-perpendicular
kinematics). We have selected the energy transfer to correspond almost to the QE peak value, where one expects the validity of the
impulse approximation to be highest. The value of the transfer momentum is fixed to $q=500$ MeV/c ($\omega =132$ MeV) 
and results are presented versus the 
missing momentum $p_m$, which in this section is written $p$ for simplicity. Finally, in most of the cases we have chosen the azimuthal angle $\phi_N$ equal to zero. Only when discussing the helicity
asymmetry do we analyze the effects linked to a selection of various $\phi_N$-values.

\subsection{Relativistic Plane-Wave Impulse Approximation (RPWIA)}

We start our study with the simple RPWIA case. 
The general formalism for the PV responses has been presented
in the previous section with explicit expressions given in Appendix~\ref{apendiceRPWIA}. 
Our discussion follows closely the analysis presented in
\cite{Caballero98a,Martinez02a} for the case of purely EM unpolarized and polarized responses. 
We show results for the six interference responses. 
First we restrict our attention to the separate $uu$, $uv$ and $vv$ single-nucleon responses ${\cal R}^\alpha_x$ analyzing off-shell and gauge ambiguities in addition to the particular contribution of each component: positive-energy, negative-energy and
the $uv$ interference term. 
Then we present results for the nuclear/hadronic responses.

\subsubsection{PV single-nucleon responses}

\begin{figure}[tbh]
    \centering
        \includegraphics[width=.78\textwidth,angle=0]{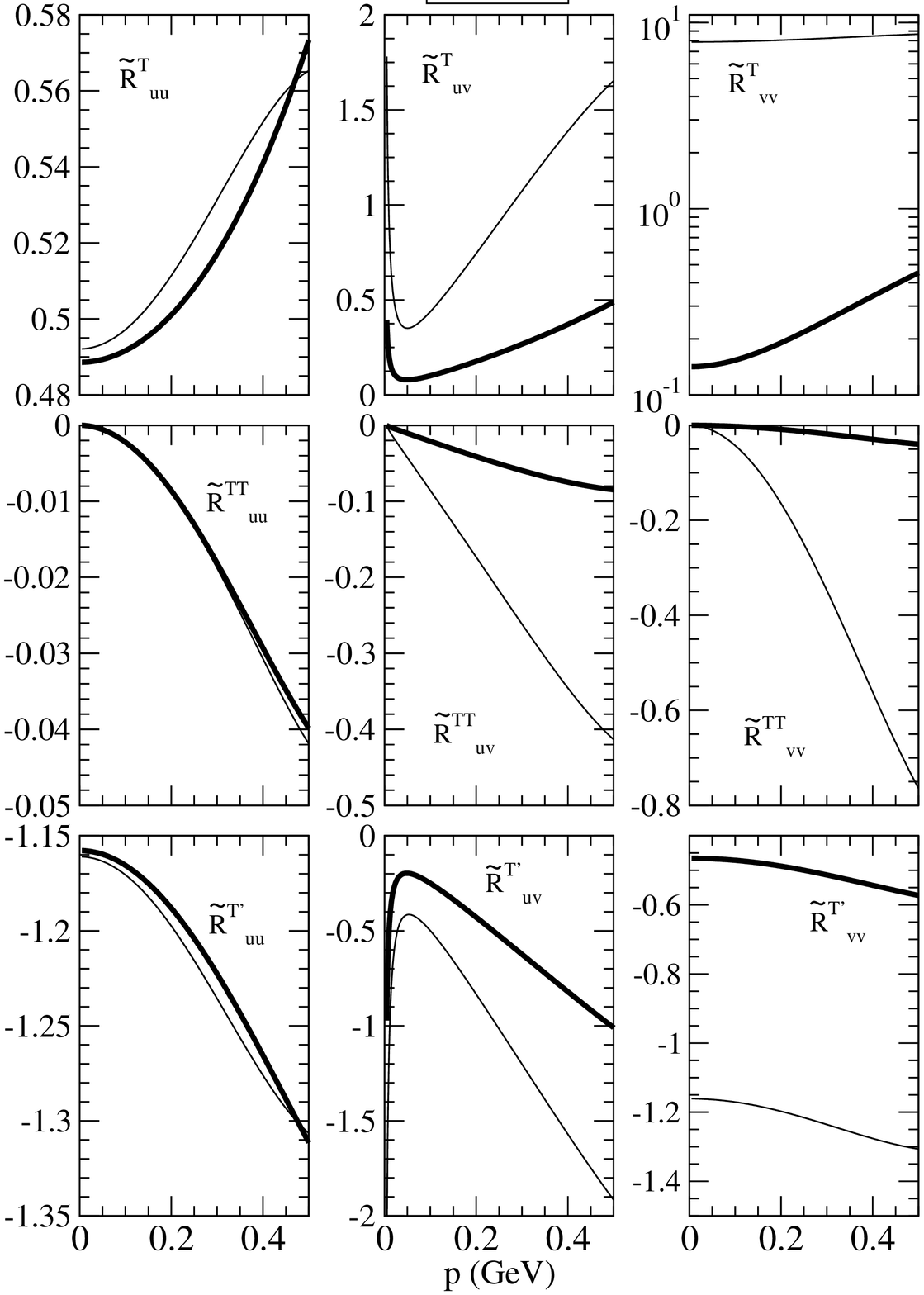}
    \caption{Transverse single-nucleon responses: $T$ (upper panels), $TT$ (middle) and  
    $T'$ (bottom). The separate projection components are shown: $uu$ (left panels),
      $uv$ (central) and $vv$ (right). Results are presented for the CC1 (thin lines) and CC2 (thick) current operators.}
    \label{fig:SN-trans}
\end{figure}
\begin{figure}[tbp]
    \centering
        \includegraphics[width=.78\textwidth,angle=0]
        {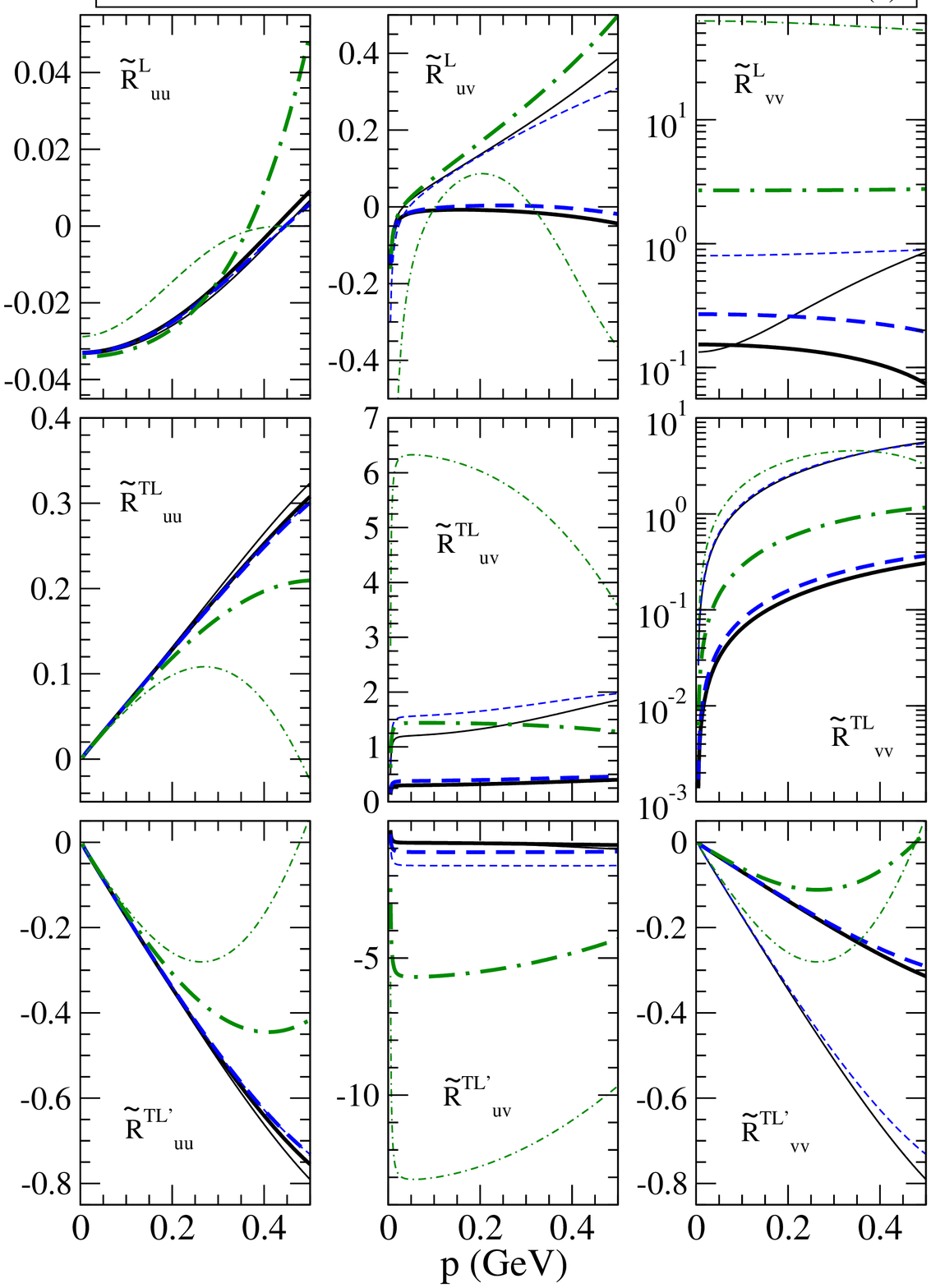}
    \caption{(Color online) As for Fig.~\ref{fig:SN-trans}, except that now the results are for the longitudinal responses: $L$ (upper panels),
      interference $TL$ (middle) and $TL'$ (bottom). Again the two currents have been selected, CC1 (thin lines) and 
      CC2 (thick), and the three gauges: Landau (solid lines), Coulomb (dashed) and Weyl (dot-dashed).}
    \label{fig:SN-long}
\end{figure}

Figures~\ref{fig:SN-trans} and \ref{fig:SN-long} show the PV single-nucleon responses. Let us start our discussion with
the case of the purely transverse channel (Fig.~\ref{fig:SN-trans}). Here, the responses depend only on the particular current
operator selected, CC1 {\it vs} CC2, but not on the gauge. Moreover, the current operator does not introduce
significant effects in the $uu$ contributions (left panels). On the contrary, the interference $uv$ and, particularly, the negative-energy
$vv$ terms are strongly affected by the current operator leading the CC1 prescription to the largest contribution (in absolute value).
Therefore, one concludes that relativistic dynamical effects do depend very much on the particular choice of the current operator.

The purely longitudinal and interference longitudinal-transverse single-nucleon responses are shown in Fig.~\ref{fig:SN-long}.
In this case, the presence of the longitudinal channel leads to differences when comparing various gauges in addition to the
particular current operator selected. Concerning the effects introduced by the operator, these are rather similar to the ones
already observed for the purely transverse responses (see previous figure). Hence we restrict our attention to the ambiguities
that emerge from the particular {\it ``gauge''} considered: Landau (NCC1 \& NCC2), Coulomb (CC1(0) \& CC2(0)) and Weyl (CC1(3) \& CC2(3)). 
As observed, results for Coulomb and Landau gauges are very similar in all cases no matter which specific current operator is selected.
On the contrary, results corresponding to the Weyl gauge (CC1(3)/CC2(3)) lead to very significant differences even in the case of the
purely positive-energy contribution $uu$.

Summarizing, the {\it ``off-shell''} effects observed in the PV single-nucleon responses are very similar to the ones
already presented for the purely EM responses~ \cite{Caballero98a}. 
Since such effects are directly linked to the vector part, 
$\sim\gamma^\mu$, in the current, the general arguments already presented in \cite{Caballero98a} also apply here. First the discrepancy
between CC1 and CC2 results are larger for the $uv$ terms, and particularly, for the $vv$ terms. On the contrary, the $uu$ contributions
show a very mild dependence on the current operator selected (see~\cite{Caballero98a} for an explanation of these effects).
Second the results corresponding to Landau and Coulomb gauges are always similar, whereas those obtained with the Weyl gauge depart significantly.
This behavior can be understood by taking the difference between the longitudinal current matrix elements,
$J^L=J^0-\frac{\omega}{q}J^3$, evaluated within the different gauges (see \cite{Caballero98a} for details).

\subsubsection{PV hadronic responses}\label{exclResp}

\begin{figure}[htbp]
    \centering
        \includegraphics[width=.6\textwidth,angle=270]
        {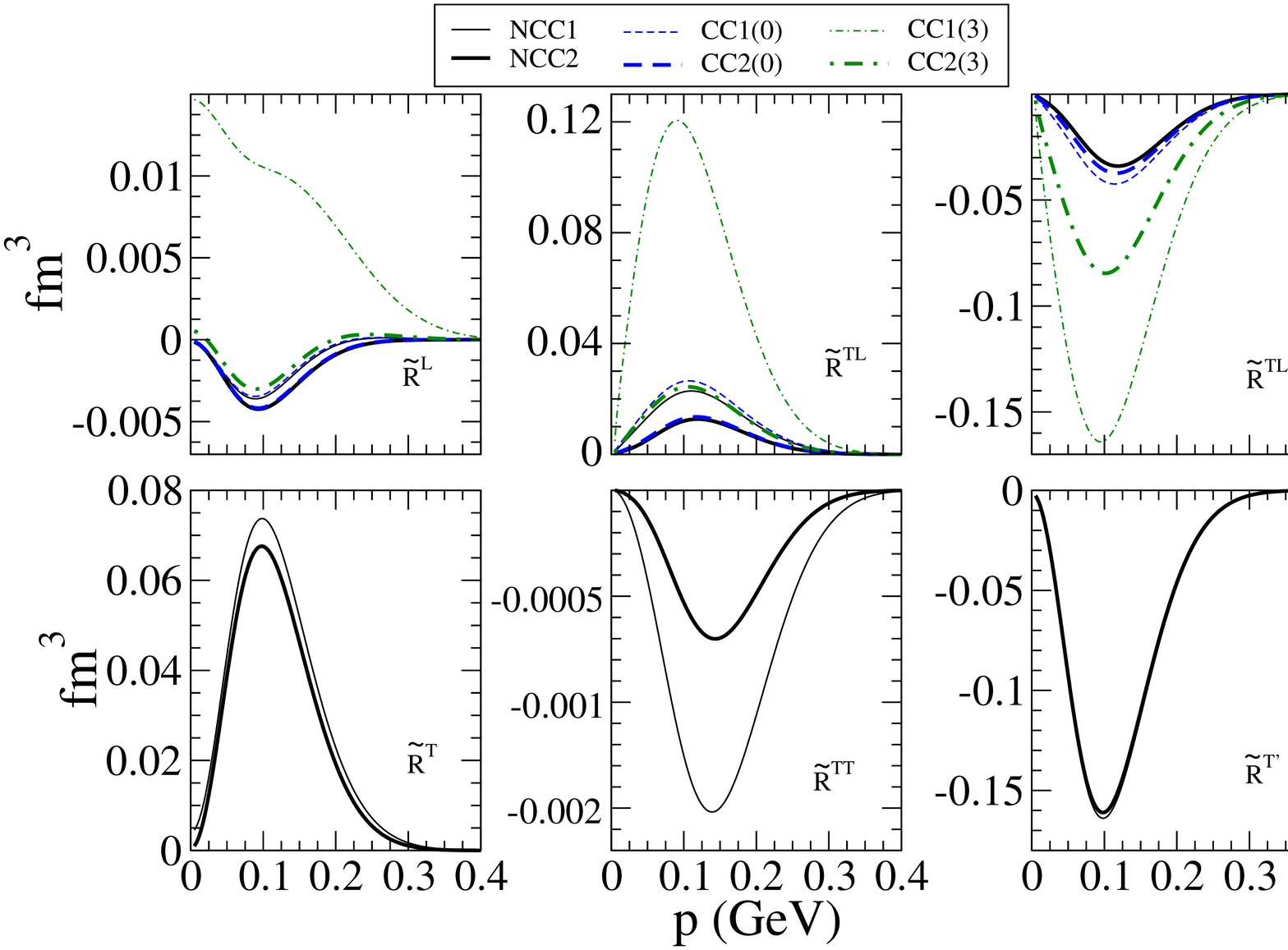}
    \caption{(Color online) Hadronic response functions for a proton in the $1p_{1/2}$-shell in $^{16}$O.
      The labels of the various curves are as in Fig.~\ref{fig:SN-long}.}
    \label{fig:excl-resp-gauge}
\end{figure}

\begin{figure}[htbp]
    \centering
        \includegraphics[width=.6\textwidth,angle=270]
        {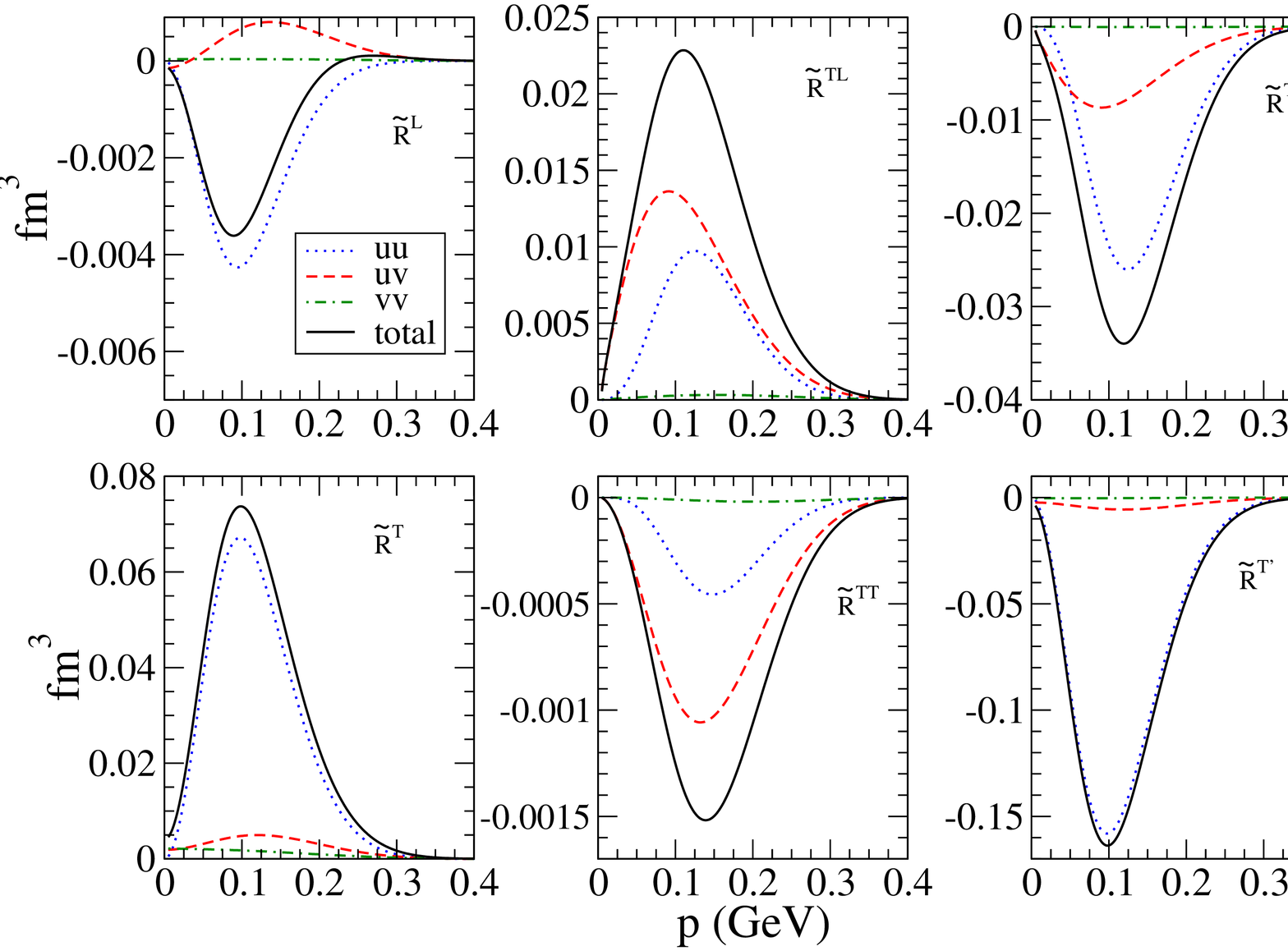}\\
      \vspace{1cm}
        \includegraphics[width=.6\textwidth,angle=270]
        {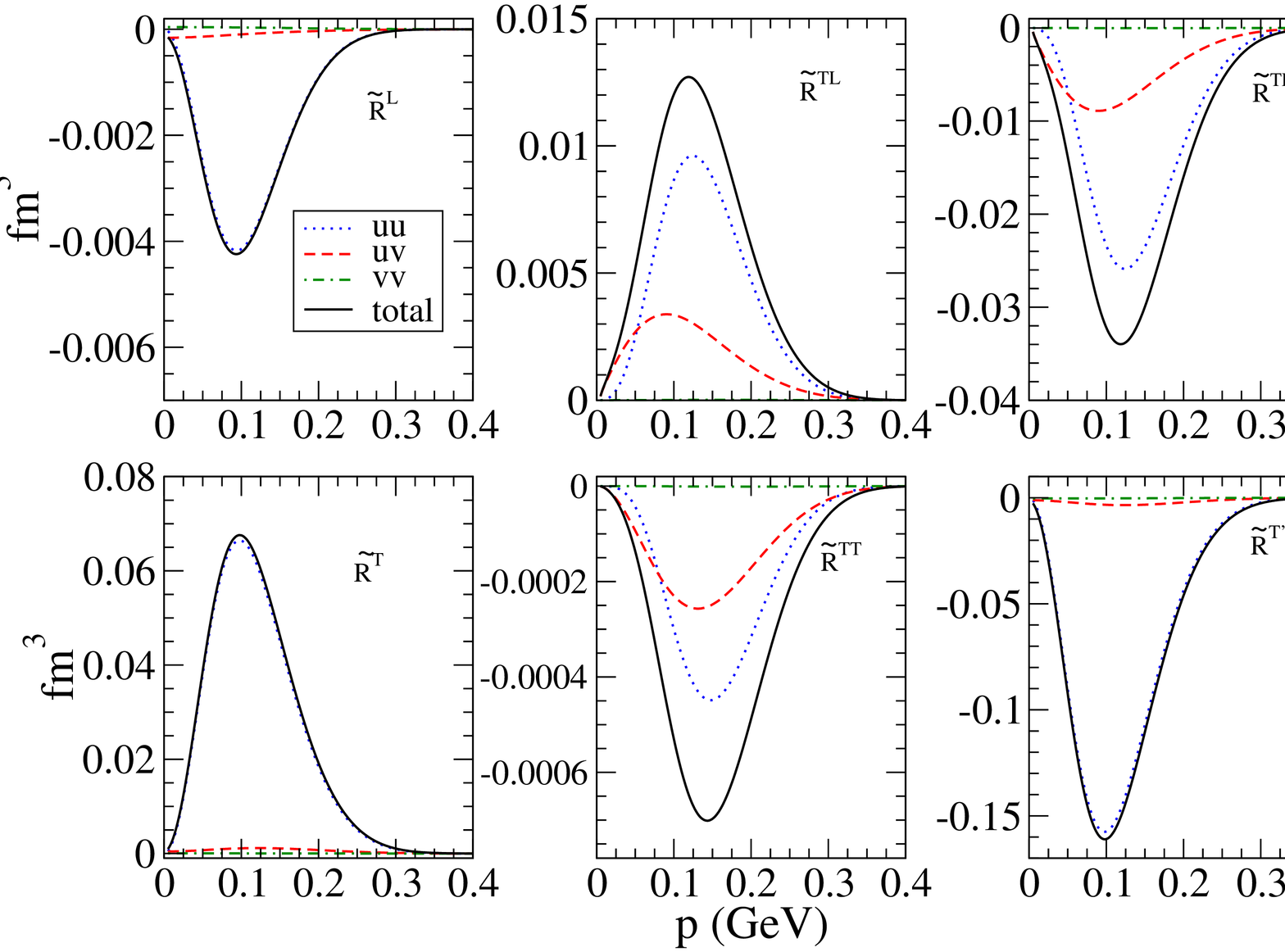}
    \caption{(Color online) Energy projection contributions to the hadronic response functions for $1p_{1/2}$-shell in $^{16}$O:
    $uu$ (blue dotted line), $uv$ (red dashed) and $vv$ (green dot-dashed). The total responses are represented by
      the solid black lines. The six top panels correspond to results obtained with the Landau gauge and CC1 current
      operator, whereas the bottom ones refer to the CC2 current.}
    \label{fig:excl-proy}
\end{figure}

In this section we present and analyze the PV hadronic responses corresponding to 
the case of protons in the $1p_{1/2}$-shell for $^{16}$O (Figs.~\ref{fig:excl-resp-gauge} and \ref{fig:excl-proy}). 
As already shown, these responses are given as products of the single-nucleon responses and the corresponding momentum
distribution components: $N_{uu}$, $N_{uv}$ and $N_{vv}$.
In Fig.~\ref{fig:excl-resp-gauge} we evaluate the effects associated with the choice of the current operator and the gauge. 
These results are consistent with the previous studies applied to the EM responses. Note that the largest discrepancies emerge with
the Weyl gauge. On the contrary, Landau and Coulomb gauges lead to rather similar results.
It is important to point out the extremely different behavior shown by the results obtained with the Weyl gauge: CC1(3) and CC2(3). 
Note the discrepancy between the CC1(3) longitudinal response and the remaining ones. This result can be ascribed to the magnitude of the
single-nucleon component $\widetilde{{\cal R}}_{vv}^L$ (see Fig.~\ref{fig:SN-long}), that is dominant for all momenta considered.

In what follows we restrict our attention to the Landau gauge, {\it i.e.,} NCC1 and NCC2 prescriptions. Results are similar within the
Coulomb gauge, whereas Weyl ones are dismissed because they fail in describing cross sections and
polarization ratios data for different kinematics~\cite{Udias01}.

In Fig.~\ref{fig:excl-proy} we show the PV hadronic responses isolating the specific contributions given by the
components $uu$, $uv$ and $vv$. As shown, the $vv$ term (green dot-dashed line) is negligible in most of the cases. 
This is a consequence of the value of the momentum distribution $N_{vv}$: one order (several) of magnitude smaller
than $N_{uv}$ ($N_{uu}$) (see \cite{Caballero98a}). 
On the contrary, the particular contribution of the interference
$uv$ component (red dashed line) depends on the specific response considered and the off-shell prescriptions selected.
As already commented on, the use of the CC1 operator tends to maximize the role played by the interference $uv$ terms.
In particular, it is noteworthy to point out the significant $uv$ contribution in the responses: 
$\widetilde{R}^{TL}$, $\widetilde{R}^{TL'}$ and $\widetilde{R}^{TT}$.
Notice that such contribution is even bigger than the purely $uu$ term in the case of the CC1 current and 
the responses $\widetilde{R}^{TL}$ and $\widetilde{R}^{TT}$ (upper central panels).
Although not shown, similar results and comments apply to the case of a neutron in the $p_{1/2}$-shell being emitted from the nucleus.
This outcome differs from the one pertained to the EM responses where the longitudinal contribution (related with the electric charge) 
is very small in the neutron responses.

\begin{figure}[htbp]
    \centering
        \includegraphics[width=.6\textwidth,angle=270]
        {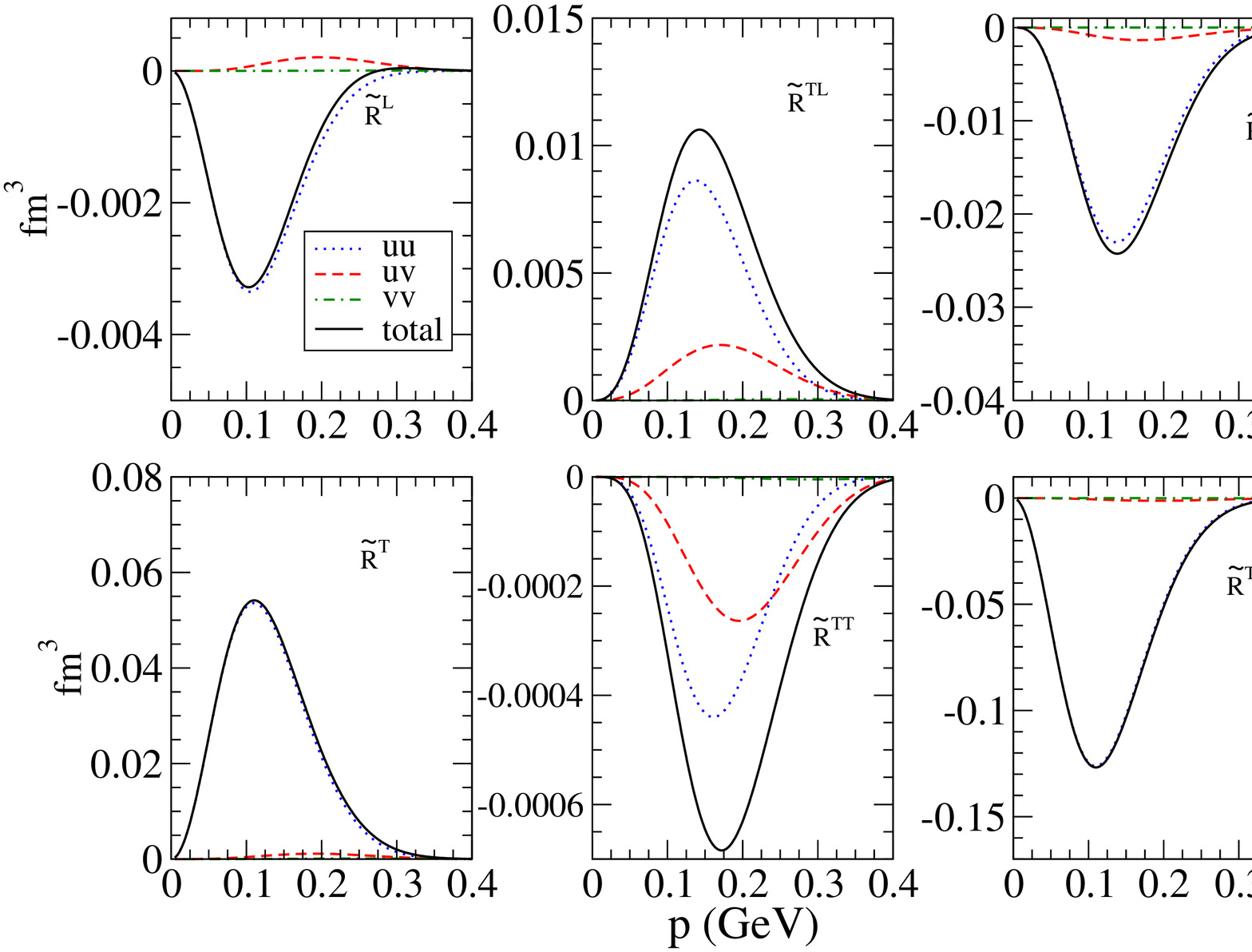}\\
      \vspace{1cm}
        \includegraphics[width=.6\textwidth,angle=270]
        {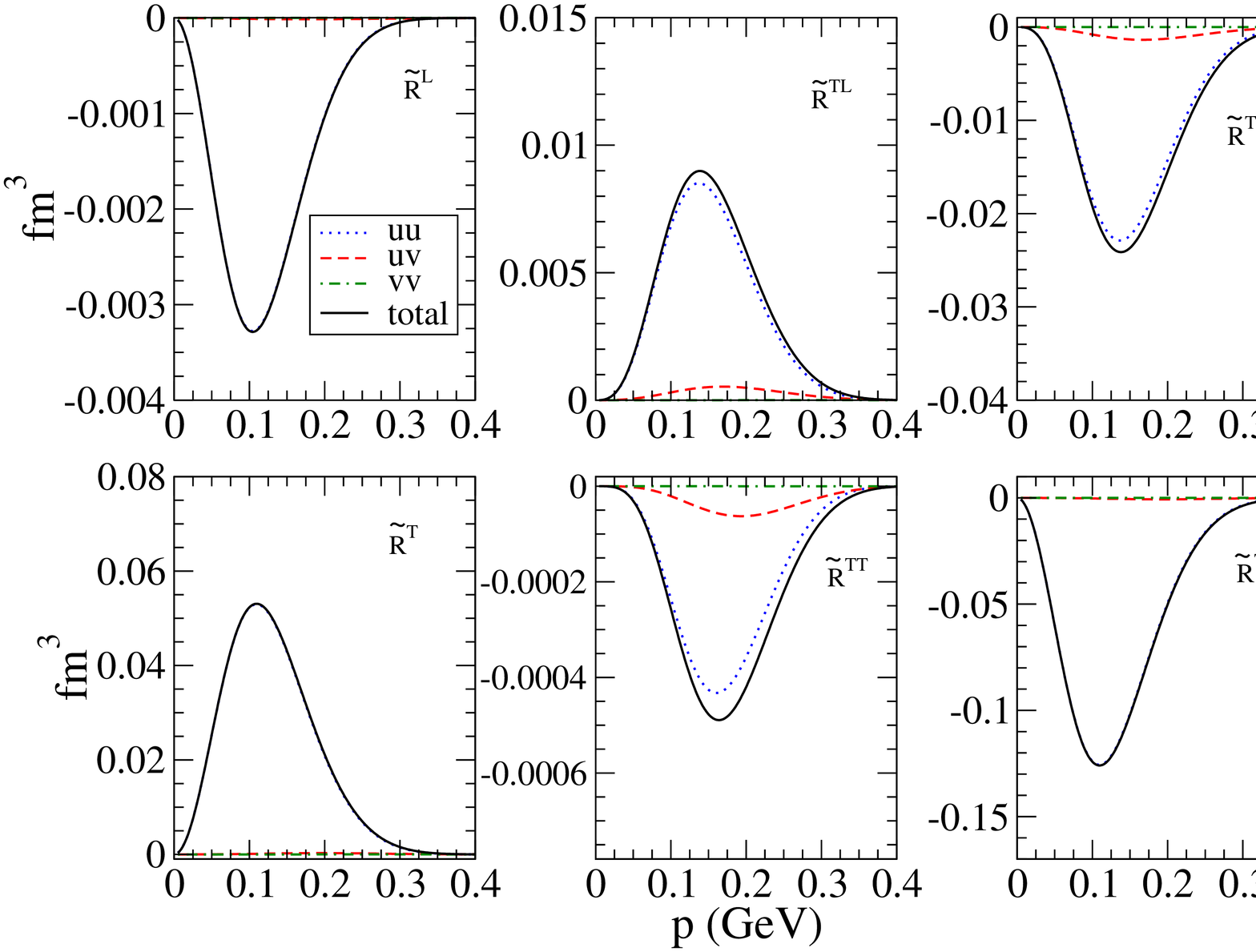}
    \caption{(Color online) As for Fig.~\ref{fig:excl-proy}, but now for a proton in the $1p_{3/2}$-shell in $^{16}$O.}
    \label{fig:excl-proy_p3p}
\end{figure}

For completeness, we also present in Fig.~\ref{fig:excl-proy_p3p} the PV hadronic responses corresponding to a
$1p_{3/2}$-shell proton in $^{16}$O. Comparing these results with the previous ones, {\it i.e.,} proton in the $1p_{1/2}$-shell 
(Fig.~\ref{fig:excl-proy}), one observes the significant reduction in the effects associated with the $uv$ (and $vv$)
components. This can be easily explained taking into account the different role played by the lower components of the
bound nucleon wave function for different spin-orbit parner shells. In fact, for the jack-knifed states ($p_{1/2}$) the
amplitudes of the negative-energy projections are much larger than those for the stretched states ($p_{3/2}$). This is
due to the different quantum number $\overline{\ell}$ of the lower components in the two kinds of states: $\overline{\ell}=0$
($\overline{\ell}=2$) for $p_{1/2}$ ($p_{3/2}$) states (see \cite{Caballero98b}).

\subsection{Final-State Interactions (FSI)}\label{excluFSI}

\begin{figure}[htbp]
    \centering
        \includegraphics[width=.6\textwidth,angle=270]
        {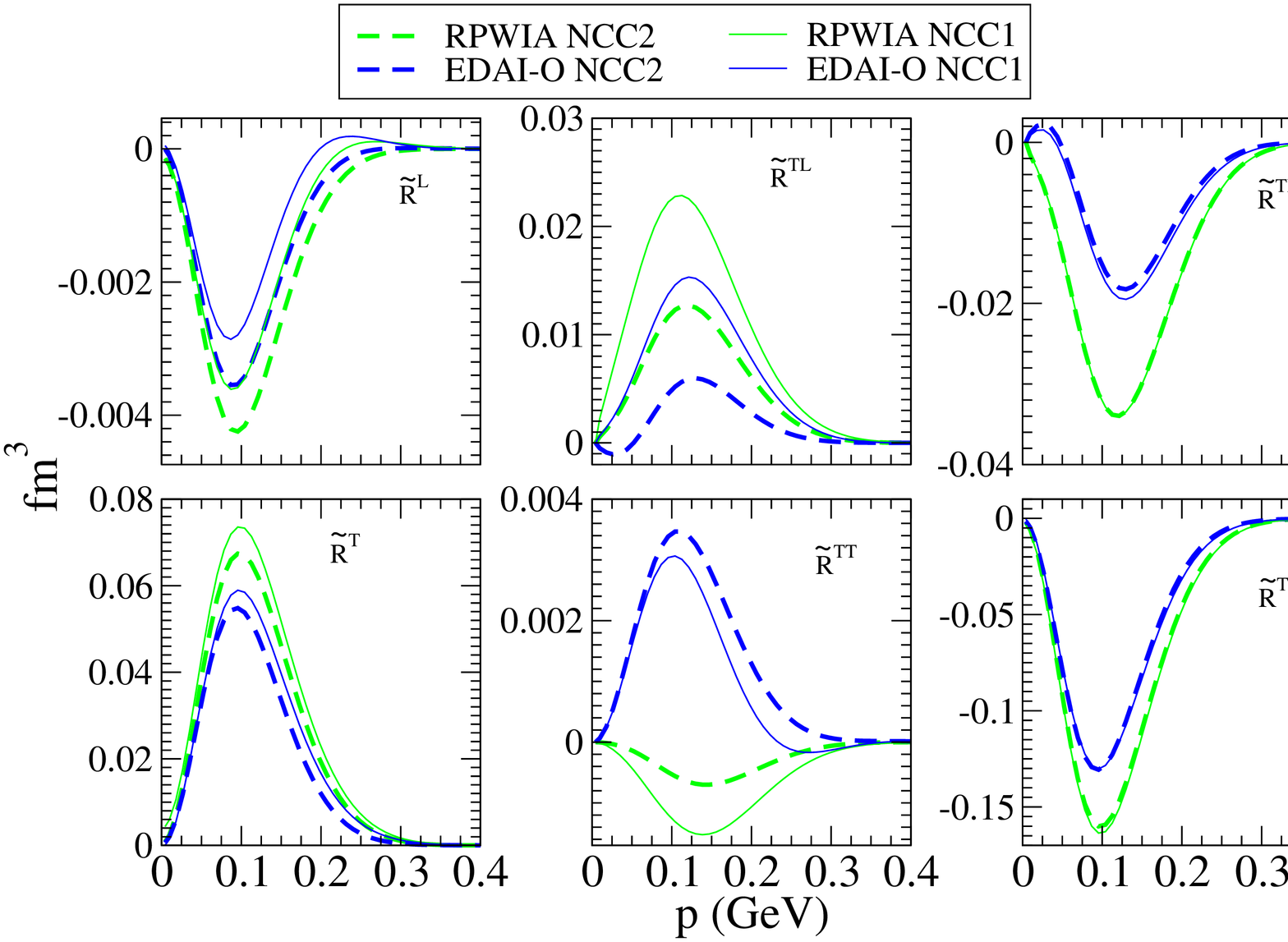}\\
        \includegraphics[width=.6\textwidth,angle=270]
        {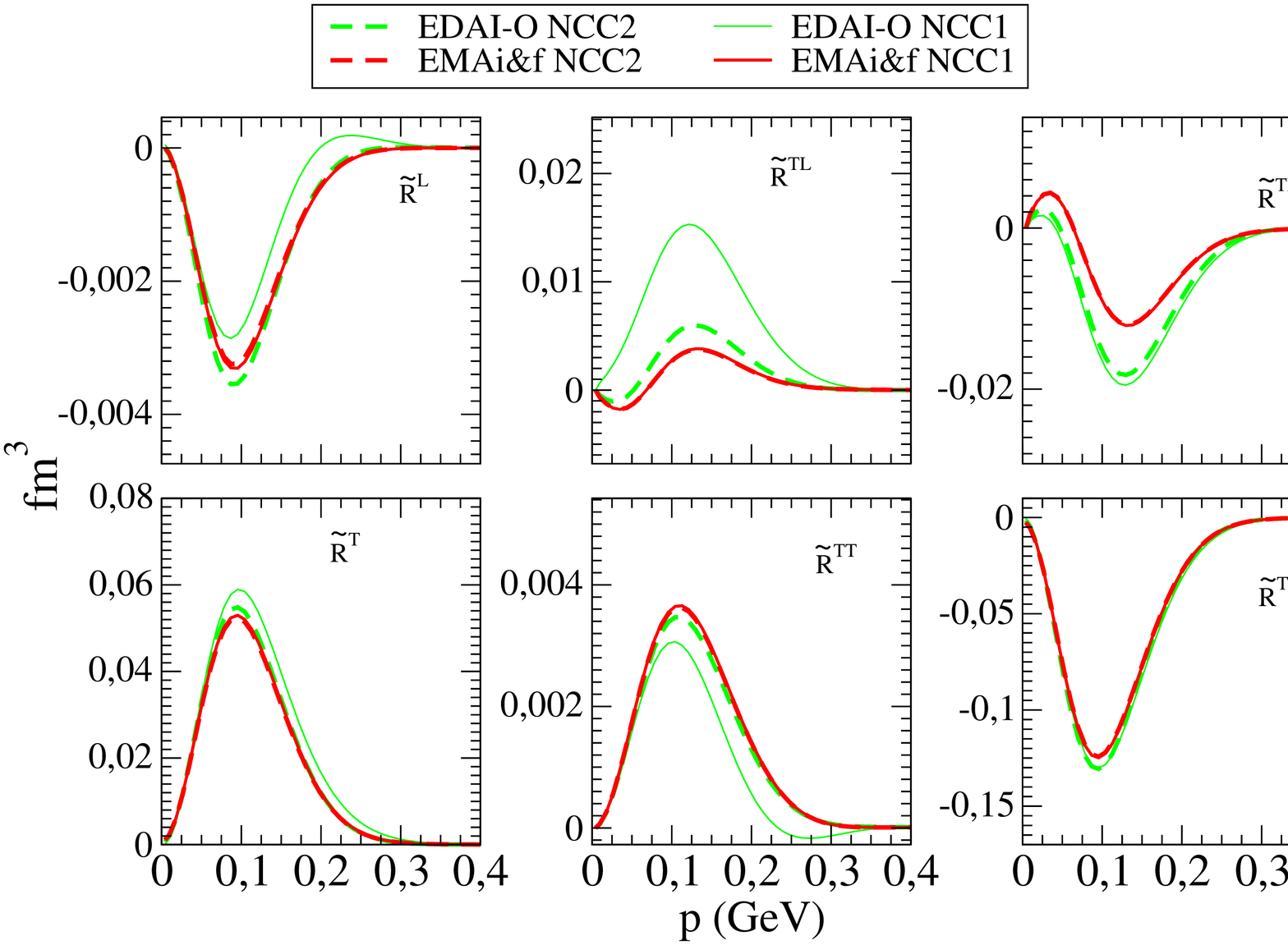}        
    \caption{(Color online) PV hadronic responses for a proton in the $1p_{1/2}$-shell in $^{16}$O. 
    In the upper panels results in RPWIA (green lines) are compared with the RDWIA ones using the EDAI-O optical potential (blue).
    The bottom panels present RDWIA-EDAI-O results (green) compared with the ones evaluated with the effective momentum approximation (EMA) (red). 
    See text for details.}
    \label{fig:excl_WNC_resp_FSI}
\end{figure}

In this section we analyze the effects introduced by the description of the FSI between the ejected nucleon
and the residual nucleus. Results for the PV exclusive responses are shown in Fig.~\ref{fig:excl_WNC_resp_FSI}.
As already mentioned in previous sections, in this work FSI are accounted for by solving the Dirac equation in the presence of 
complex phenomenological optical potentials fitted to elastic nucleon scattering data. 
In particular, here we restrict ourselves to the use of the energy-dependent $A$-independent EDAI-O potential. 
The use of other potentials does not modify the main conclusions.
The kinematics have been fixed as in the previous figures, namely,
the transfer momentum is fixed to $q=500$ MeV/c and the energy transfer is chosen to be in the maximum of the QE 
peak, $\omega=132$ MeV. Coplanar kinematics, {\it i.e.,} $\phi_N=0$, have been selected.

All results presented in Fig.~\ref{fig:excl_WNC_resp_FSI} correspond to the case of a proton being ejected from the $1p_{1/2}$-shell in $^{16}$O. 
In the six upper panels we compare the results evaluated in the RPWIA limit (green lines) with the ones obtained by including FSI (blue lines). 
As observed, in most of the cases FSI lead to a significant reduction in
the magnitude of the corresponding response, being larger for the longitudinal-transverse interference responses:
$\widetilde{R}^{TL}$ and $\widetilde{R}^{TL'}$ (they are reduced by a factor of two). 
A particular comment applies to
$\widetilde{R}^{TT}$: FSI effects completely modify the behavior shown by the response, even changing the global sign (from negative RPWIA values 
to positive RDWIA ones). However, note the smallness of $\widetilde{R}^{TT}$ 
(likewise for $\widetilde{R}^{L}$) whose contribution is negligible compared with the remaining responses. 
Hence it is not strange that this response shows a very high sensitivity to FSI and its
particular description. 
Finally, concerning the differences ascribed to the use of a particular current operator, {\it i.e.,} CC1 {\it vs} CC2, the results in Fig.~\ref{fig:excl_WNC_resp_FSI} 
show a similar behavior for the two approaches, {\it viz.} RPWIA and RDWIA.

To conclude, we analyze the so-called dynamical relativistic effects, that is, relativistic effects associated with the description of the nucleon wave functions. 
We compare our fully relativistic RDWIA results with those obtained by projecting out the negative-energy components in both the bound and scattered nucleon 
wave functions. 
This is equivalent to the $uu$, $uv$ and $vv$ decomposition shown for the RPWIA case in the previous section. Here we apply the study to the distorted calculation. 
In the six bottom panels in Fig.~\ref{fig:excl_WNC_resp_FSI} we compare the RDWIA results with the positive-energy projected ones. 
The latter approach is simply known as effective momentum approximation (EMA) (see
\cite{Kelly97,Kelly99,Udias95,Udias01,Vignote04} for details on how the EMA approach is defined).
As expected, the differences introduced by the choice of the current operator and/or gauge are much smaller within the EMA limit.
Note that EMA does not incorporate contributions linked to the $uv$ or $vv$ terms; hence the difference in the responses come solely from the ``off-shell'' nucleon effects. 
This behavior, shown in Fig.~\ref{fig:excl_WNC_resp_FSI} for the PV responses,
was already observed for the purely EM ones~\cite{Caballero98a}.

\subsection{Exclusive Helicity Asymmetry}\label{AsiExcl}

We can introduce an asymmetry for the exclusive process $A(\vec{e},e'N)B$, defined as the ratio between the difference and the
sum of {\it exclusive} cross sections evaluated for positive and negative electron helicity, respectively. This is given as
\be
 {\cal A}_{excl} = \frac{\sigma^{+}-\sigma^{-}}{\sigma^{+}+\sigma^{-}}\, ,
\label{asymmetry-excl}
\ee
where $\sigma^{+/-}$ represents the differential cross section corresponding to positive/negative incident electron helicity~(\ref{sigexclPW}). 
However, there exists a crucial difference between the above asymmetry defined for $(\vec{e},e'N)$  
processes and
the corresponding one constructed for inclusive $(\vec{e},e')$  
reactions (likewise, for elastic PV electron scattering on the proton).
Whereas for $(\vec{e},e')$  and elastic $PVep$ scattering the asymmetry 
is only different from zero because of the role played by the weak interaction, in the case of $(\vec{e},e'N)$  
processes the purely
EM interaction also gives a contribution to $\sigma^{+}-\sigma^{-}$ through the fifth EM response function (see discussion of the 
previous figures). Therefore, the {\it exclusive} helicity asymmetry introduced in this section is not, in principle, a good observable
to analyze effects linked to the weak interaction; the purely EM one dominates by orders of magnitude. On the contrary, this observable
is particularly suited to study FSI. In what follows we discuss in detail these results analyzing
under which conditions both the purely EM and the WNC interference contributions give similar results. 
In other words, we discuss
the limits under which the study of PV responses makes sense for $A(\vec{e},e'N)B$ reactions.

From the general expression for the cross section in Eq.~(\ref{sigexclR1}) we can isolate in the helicity asymmetry in Eq.~(\ref{asymmetry-excl})
its purely EM contribution, ${\cal A}_{excl}^{EM}$, and the one associated with the presence of the weak interaction, 
${\cal A}_{excl}^{WNC}$:
\be
 {\cal A}_{excl} (\theta_e,q,\omega,E_m,\phi_N,p) 
          = {\cal A}_{excl}^{EM}(\theta_e,q,\omega,E_m,\phi_N,p) 
          + {\cal A}_{excl}^{WNC}(\theta_e,q,\omega,E_m,\phi_N,p)\, .
          \label{Aexcl}
\ee
Here we express the explicit dependence of the asymmetry with all the kinematical variables. In terms of the nuclear
response functions, the interference ${\cal A}_{excl}^{WNC}$ term can be written in the form:
\begin{eqnarray}
 {\cal A}_{excl}^{WNC} 
          &=& \frac{{\cal A}_0}{2 {\cal G}^2} \left[
          a_A\left(v_L \widetilde{R}^L + v_T \widetilde{R}^T 
          + v_{TT} \widetilde{R}^{TT} + v_{TL}\widetilde{R}^{TL}\right)
          \right.\nonumber\\
 %%%%
          &-& \left. a_V \left(v_{T'} \widetilde{R}^{T'} + v_{TL'} 
          \widetilde{R}^{TL'}\right)
           \right]\, ,\label{AexclWNC}
\end{eqnarray}
where we have introduced the function ${\cal G}^2\approx v_L R^L + v_T R^T + v_{TT} R^{TT} + v_{TL} R^{TL}$~\footnote{We have neglected 
in ${\cal G}^2$ the very small contribution given by the PV responses.}.

The purely EM contribution, ${\cal A}_{excl}^{EM}$, is simply given by the fifth response function $R^{TL'}$:
\be
{\cal A}_{excl}^{EM}=\frac{v_{TL'}R^{TL'}}{{\cal G}^2}\,.\label{AexclEM}
\ee
Note that $R^{TL'}$ only enters when FSI are incorporated in the analysis ($R^{TL'}=0$ in RPWIA). Moreover, the dependence on the
azimuthal angle $\phi_N$ is simply given through $\sin\phi_N$, {\it i.e.,} in the limit of coplanar kinematics, $\phi_N=0^o$, $180^o$, the
helicity asymmetry being different from zero is solely due to the weak interaction. 

In what follows we present a brief analysis of the helicity asymmetry showing the results obtained for different kinematics and 
evaluating the role of FSI. We also analyze the effects associated with the lower components in the relativistic wave functions and
with the particular choice of the nucleon current operator. Our interest is to determine under which conditions the 
{\it ``exclusive''} helicity asymmetry can be appropriate to get information on the PV response functions
 
\subsubsection{Coplanar kinematics: ${\bs\phi_N=0}$} 

\begin{figure}[ht]
    \centering
        \includegraphics[width=.55\textwidth,angle=270]
        {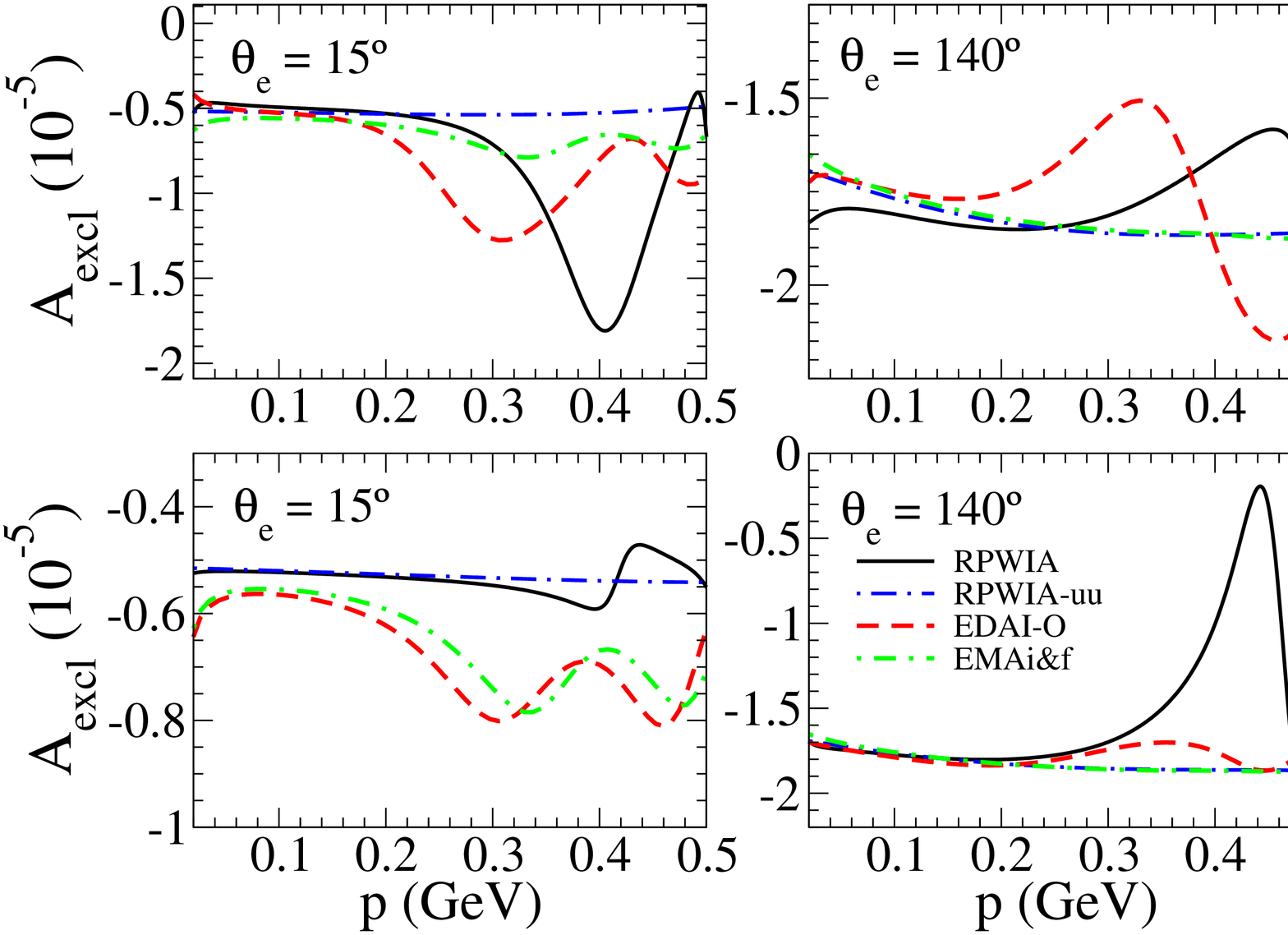}
    \caption{(Color online) Asymmetry ${\cal A}_{excl}$ for the case of a proton in the $1p_{1/2}$-shell in $^{16}$O. 
    The kinematics are fixed to $q=0.5$ GeV/c and $\omega=0.132$ GeV, and two values for the scattering angle have been selected:
    $\theta_e=15^o$ (forward scattering, left panels) and $\theta_e=140^o$ (backward, right panels). Coplanar kinematics are used,
    {\it i.e.,} $\phi_N=0^o$. The top (bottom) panels refer to results evaluated with the NCC1 (NCC2) prescriptions. In each case we compare
    the asymmetry corresponding to the RPWIA, RPWIA-uu, RDWIA and EMA approaches. FSI have been evaluated using the EDAIO potential.}
    \label{fig:exclu_asym_PWvsDW}
\end{figure}

For coplanar kinematics the helicity asymmetry reduces to ${\cal A}_{excl}={\cal A}_{excl}^{WNC}$.
In Fig.~\ref{fig:exclu_asym_PWvsDW} we present the asymmetry corresponding to the case of protons in the $1p_{1/2}$-shell in
$^{16}$O. As in previous sections, the kinematics have been fixed to $q=500$ MeV, $\omega=132$ MeV, $\phi_N=0$ and two values for the
scattering angle: $\theta_e=15^o$ (forward scattering; left panels), and $\theta_e=140^o$ (backward angles; right panels). 
The top (bottom) panels show results for the Landau gauge and the CC1 (CC2) current.

RPWIA results are shown with black lines (fully relativistic responses) and blue lines (positive-energy projection). The comparison
between the two approaches clearly shows the role played by the lower components in the bound nucleon wave function. In the region of 
the missing momentum where the responses attain their maximum values, {\it i.e.,} $50 < p < 200$ MeV, the asymmetry does show a tiny sensitivity with
the momentum, leading both approaches, RPWIA and projected, to rather similar results. On the contrary, for higher values,
$p\geq 250$ MeV, the relative contribution of the lower (negative-energy) components starts to increase, causing significant departures in the various curves. This result applies for both of the scattering angles selected.

Concerning the effects linked to the choice of the current operator, {\it i.e.,} NCC1 {\it vs} NCC2, the differences are negligible within
the projected-energy approach. However, these discrepancies get higher when the contribution coming from the lower-energy components
is included. This is particularly true for large missing momenta and it is basically given by the interference $uv$ terms. Notice that the 
discrepancy between the curves can reach a factor $\sim 4$ in the case of forward kinematics, and $\sim 2$ at backward. Moreover, 
the lower components are responsible of the oscillating behavior shown by the asymmetry.

Figure~\ref{fig:exclu_asym_PWvsDW} also contains the results obtained with FSI. Here we distinguish the fully relativistic
distorted (RDWIA) calculation (red lines) from the effective momentum approximation (EMA) (green lines), 
{\it i.e.,} projecting over positive-energy
components both the bound and the scattered nucleon wave functions. From the comparison between RPWIA, RDWIA and EMA results we
observe that at $p\sim 0.1$ MeV (maxima in the responses) the three approaches lead to very similar results. Although not shown,
this comment also applies to results obtained with other gauges. To conclude, let us note the important role played by FSI and its
dependence on the particular description used to describe such effects: see the comparison between RDWIA and EMA. The most sensitive region
occurs at high missing momenta (the region where the responses, and likewise the cross section, are very small).

The general conclusions reached from results in Fig.~\ref{fig:exclu_asym_PWvsDW} can be extended to other values of the transferred
momentum and energy close to the QE peak.

\subsubsection{Non-coplanar kinematics: ${\bs\phi_N\neq0}$}\label{Aexclneq0}

As already mentioned in previous sections, the so-called fifth, purely EM response $R^{TL'}$ only enters in the analysis of $(\vec{e},e'N)$
reactions when FSI are incorporated. Moreover, this response contributes only for non-coplanar kinematics because its dependence
with the azimuthal angle is simply given through $\sin\phi_N$. The determination of the fifth response is also linked to the measurement
of the incident electron helicity. Therefore, its contribution to the helicity asymmetry can be very relevant, particularly, much
more important than the contribution coming from the PV responses.

This problem is considered in the discussion that follows. We consider different kinematical regimes and analyze the impact that the 
fifth EM response may have in the asymmetry, comparing its particular contribution with the one ascribed to the PV
responses. The purely EM responses, included $R^{TL'}$, are several orders of magnitude bigger than the PV ones. 
Hence we have considered specific situations, very close to the strictly coplanar kinematics, where both the purely EM and the WNC sectors may
lead to similar contributions to the helicity asymmetry. The interest in this study is to determine what level of precision should be
required on the azimuthal angle $\phi_N$ in order to isolate in the asymmetry the particular contribution associated with the PV responses. 
To simplify the discussion, in what follows all results have been obtained with the prescription NCC2.
Similar conclusions are drawn from the use of NCC1. 

\begin{figure}[htbp]
    \centering
        \includegraphics[width=.405\textwidth,angle=270]
        {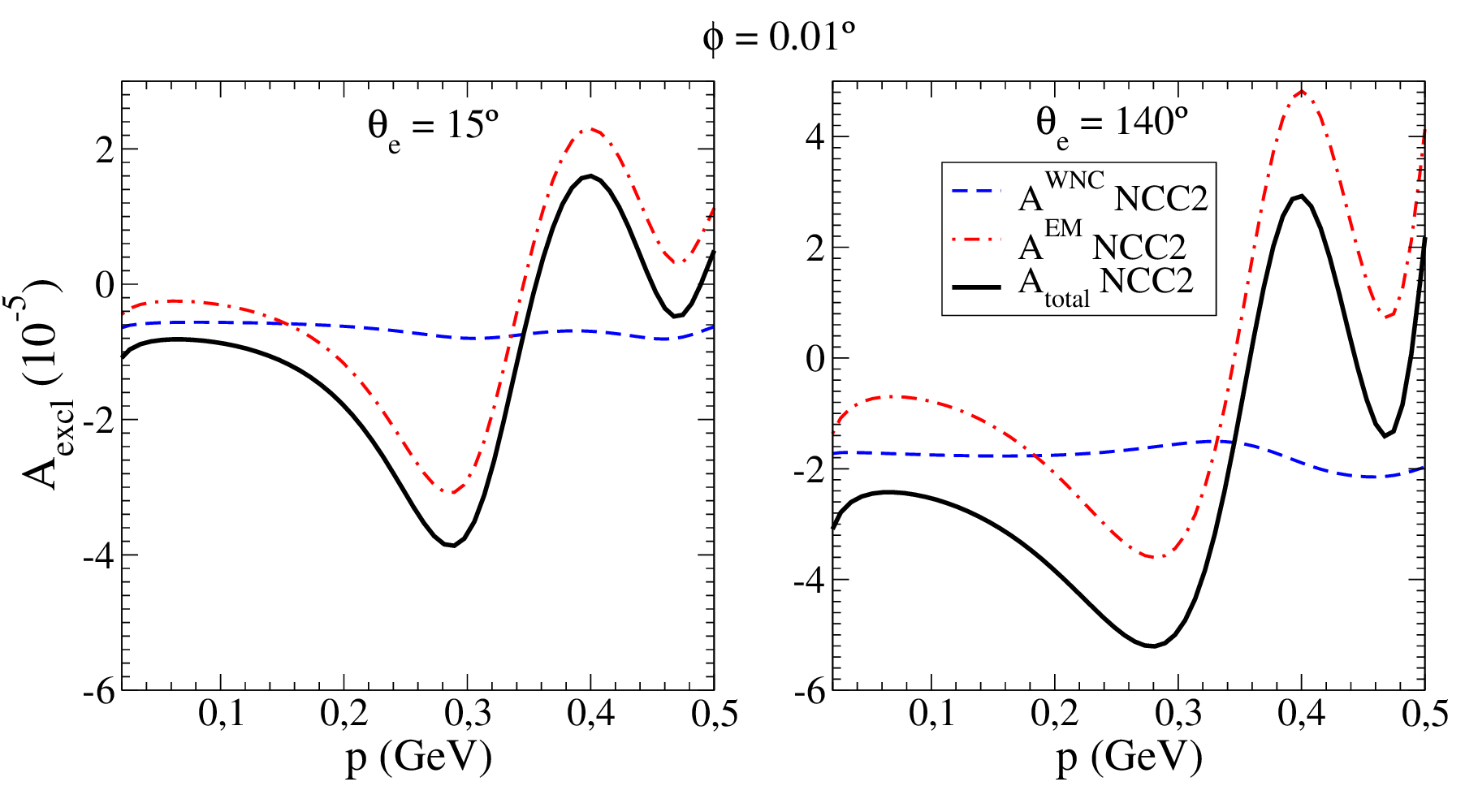}\\
        \includegraphics[width=.405\textwidth,angle=270]
        {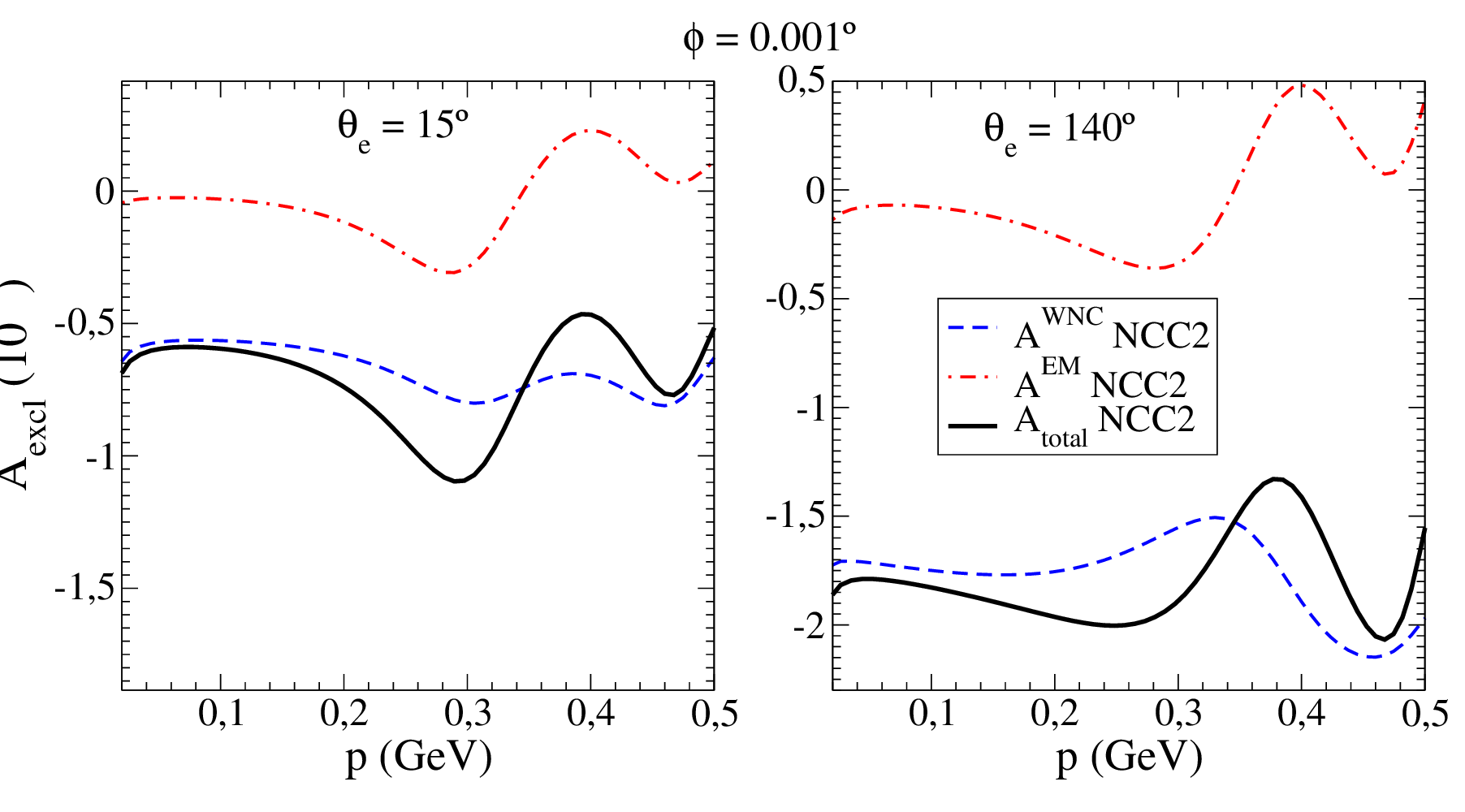}        
    \caption{(Color online) Exclusive helicity asymmetry for the $p_{1/2}$-shell (protons) in $^{16}$O. Results correspond to RDWIA and NCC2 prescription.
     The azimuthal angle has been fixed to $\phi_N=0.01^o$ (top panels) and $\phi_N=0.001^o$ (bottom). 
     Forward and backward kinematics have been explored in the left and right panels, respectively.}
    \label{fig:asy_WNCvsEM_0.0}
\end{figure}

In the top panels of Fig.~\ref{fig:asy_WNCvsEM_0.0} we show ${\cal A}_{excl}$ (black solid line) at $\phi_N=0.01^o$ and the separate contributions: 
${\cal A}_{excl}^{EM}$ (red dashed-dotted line) and ${\cal A}_{excl}^{WNC}$ (blue dashed line). 
This situation is rather close to coplanar kinematics and, 
as shown in the figure, the purely EM contribution clearly dominates both at forward (left panel) and backward (right) kinematics. 
The bottom panels show the results corresponding to an even smaller azimuthal angle, $\phi_N=0.001^o$. 
In this case, because of the $\sin\phi_N$ dependence in the fifth response, the PV contribution is dominant, and here the purely 
EM one is roughly one order of magnitude smaller. 
It is interesting to point out that, contrary to ${\cal A}^{EM}_{excl}$, the interference contribution
${\cal A}^{WNC}_{excl}$ does not show sensitivity with $\phi_N$ (in the case of very small $\phi_N$-values). On the other hand,
${\cal A}^{EM}_{excl}$ presents a much stronger sensitivity with the missing momentum, being responsible of the general 
{\it ``oscillating''} behavior shown by the helicity asymmetry. 

From this general analysis one concludes that the measurement of the {\it ``exclusive''} helicity asymmetry cannot provide information
on the PV responses unless the kinematics can be fixed with an azimuthal angle of the order of one thousandth of degree or
less. In any other situation, the asymmetry only shows effects associated with the purely EM responses.

\section{Summary and conclusions}\label{conclusionsIII}

In this work we have studied PV effects in exclusive $A(\vec{e},e'N)B$ processes in the QE regime.
Our main interest has been to explore new observables that may allow us to get new information on the nucleon structure.
In particular, PVQE reactions on complex nuclei can provide information on the WNC form factors that 
complements the one obtained from other processes such as elastic scattering off protons, elastic and QE electron 
scattering off helium~\cite{HAPPEXa,HAPPEXHe} and deuterium~\cite{SAMPLE05}, neutrino scattering, {\it etc.}

In this paper we have focused on the exclusive scattering process, $A(\vec{e},e'N)B$, in which both the scattered electron and 
the ejected nucleon are detected in coincidence. Although being aware of the sifnificant difficulties in getting information on
PV responses due to the presence of the so-called fifth EM response function, its general study can be considered as a first step
in the analysis of PVQE inclusive $(\vec{e},e')$ processes. Moreover, it can also provide additional information
associated with the off-shell properties of the nucleons and dynamical relativistic effects, that can be of great interest
in the discussion of PV $(\vec{e},e')$  observables. 
Therefore, in Sect.~\ref{SEDchapter} we presented in detail the kinematics 
and general formalism needed to compute the exclusive differential cross section and its decomposition into response functions. 
This section also contains a careful discussion of the impulse approximation (IA) and the particular models we have considered. 
Finally, a complete analytical calculation of the PV nuclear tensors (likewise for the responses) within the 
relativistic plane-wave impulse approximation (RPWIA) is presented in Sect.~\ref{RPWIAsect}.

In Sect.~\ref{resultados-excl} the results corresponding to the exclusive observables are presented and analyzed. 
First, off-shell effects in the interference responses are considered within the framework of the RPWIA. 
The single-nucleon interference responses are evaluated by isolating the different energy-projection contributions: 
$uu$, $uv$ and $vv$. Moreover, the two usual prescriptions for the nucleon (vector) current, {\it i.e.,} CC1 and CC2, have been used, 
and for the longitudinal channel, results are shown for the three gauges: Landau, Coulomb and Weyl. 
This analysis has been extended to the hadronic responses that are given as the product of the single-nucleon responses and 
the momentum distributions of the nucleon. Most of our conclusions concerning off-shell effects are consistent with those 
reported in~\cite{Caballero98a} for the EM responses.
We may summarize our main findings as follows:
\begin{itemize}
 \item The use of the CC1 current tends to magnify relativistic dynamical effects, {\it i.e.,} the contribution linked to the 
   lower components in the nucleon wave function.
   Differences between CC1 and CC2 results come from the $vv$ and, particularly, from the $uv$ contributions. 
   Notice that the momentum distribution $N_{vv}$ is several orders of magnitude smaller than $N_{uv}$.
%%%
 \item Responses evaluated with the Landau and Coulomb gauges are very similar. On the contrary, the Weyl gauge leads to 
   very significant differences that, for some kinematics, are not consistent with data for the EM responses. 
   Therefore, most of the results shown in this work correspond to the Landau gauge.
%%%
\end{itemize}
We have analyzed the effects introduced by final-state interactions (FSI) in the PV responses and have computed 
the helicity asymmetry associated with the exclusive process. The analysis of the weak interaction through PV electron 
scattering requires observables whose existence should be unequivocally linked to such interactions. This is the case of 
the helicity asymmetry defined for elastic PV electron-proton scattering as well as for QE $(\vec{e},e')$ reactions.
However, the situation is more delicate in the case of coincidence $(\vec{e},e'N)$ processes. Here
FSI give rise to the so-called fifth EM response, $R^{TL'}$, that enters in the analysis of the process when the 
incident electron helicity is measured. Moreover, its angular dependence is given through $\sin\phi_N$, and hence it only appears for
out-of-plane (non-coplanar) kinematics. This means that, unless the angle $\phi_N$ is very close to $0$ or $\pi$ 
(to an accuracy higher than one thousandth of a degree), the helicity asymmetry is completely dominated by the EM interaction; 
hence, barring such extreme circumstances, it
cannot provide information on the PV responses associated with the weak interaction.

\section*{Acknowledgements}

This work was partially supported by DGI (Spain): FIS2011-28738-C02-01, 
by the Junta de Andaluc\'ia (FQM-160) and
the Spanish Consolider-Ingenio 2000 programmed CPAN, 
and in part (TWD) by US Department of Energy under grant Contract Number
DE-FG02-94ER40818.
R.G.J. acknowledges financial help from VPPI-US (Universidad de Sevilla) 
and from the Interuniversity Attraction Poles Programme initiated 
by the Belgian Science Policy Office.

\appendix

\section{Relativistic Plane-Wave Impulse Approximation (RPWIA)}
\label{apendiceRPWIA}

In this appendix we show in detail the calculation of the hadronic tensors and responses that enter in the 
analysis of $(\vec{e},e'N)$  reactions within the 
Relativistic Plane-Wave Impulse Approximation (RPWIA). The general procedure was originally developed 
in~\cite{Caballero98a} in the case of unpolarized
EM responses, and later extended to the study of EM polarized observables~\cite{Martinez02a,Martinez02b}. 
Here we follow a similar procedure and we
apply the general formalism to the analysis of PV electron scattering. 
We isolate the contribution 
ascribed to the positive- and negative-energy
components in the bound nucleon wave function and show results for the two usual prescriptions of the current operator: CC1 and CC2.

Within RPWIA the hadronic tensors can be given in the general form:
\begin{eqnarray}
 W^{\mu\nu}&=&
 \overline{\sum_{I}}\sum_{F}
 \bigl[J_{EM}^{\mu}\bigr]^*\bigl[J_{EM}^{\nu}\bigr]\nonumber\\
&=& 
e^2\overline{\sum_{m_j}}\sum_{s_N}
\bigl[\overline{u}({\bf p}_N,s_N)\hat{J}_{EM}^{\mu}\ 
\Phi_k^{m_j}({\bf p})\bigr]^*\bigl[\overline{u}({\bf p}_N,s_N)\hat{J}_{EM}^{\nu}\ 
\Phi_k^{m_j}({\bf p})\bigr]
\end{eqnarray}
\begin{eqnarray}
 \widetilde{W}^{\mu\nu}&=&
 \overline{\sum_{I}}\sum_{F}\bigl[J_{EM}^{\mu}\bigr]^*
 \bigl[J_{WNC}^{\nu}\bigr]\nonumber\\
&=& 
\frac{eg}{4\cos\theta_W}\overline{\sum_{m_j}}\sum_{s_N}
\bigl[\overline{u}({\bf p}_N,s_N)\hat{J}_{EM}^{\mu}\ \Phi_k^{m_j}({\bf p})\bigr]^*
\bigl[\overline{u}({\bf p}_N,s_N)\hat{J}_{WNC}^{\nu}\ \Phi_k^{m_j}({\bf p})\bigr] \, .
\end{eqnarray}

The current matrix elements can be decomposed by using the completness relation:
\begin{eqnarray}
 (J^{\mu})_{u}=\sum_S\overline{u}({\bf p}_N,s_N)\hat{J}^{\mu}u({\bf p},S)\ 
 \bigl[\overline{u}({\bf p},S)\Phi_k^{m_j}({\bf p})\bigr]\, ,\\
 (J^{\mu})_{v}=\sum_S\overline{u}({\bf p}_N,s_N)\hat{J}^{\mu}v({\bf p},S)\ 
 \bigl[\overline{v}({\bf p},S)\Phi_k^{m_j}({\bf p})\bigr]\, ,
\end{eqnarray}
where the first term, labelled with the index $u$, comes from the coupling of the bound nucleon wave function with the positive-energy
Dirac spinors $u({\bf p},S)$, and the second one, $v$, is linked to the negative-energy Dirac spinors $v({\bf p},S)$.

Proceeding in this way the purely EM and PV hadronic tensors can be separated into three terms:
$W^{\mu\nu}=W^{\mu\nu}_P+W^{\mu\nu}_C+W^{\mu\nu}_N$, where $W^{\mu\nu}_P$ ($W^{\mu\nu}_N$) is the contribution from positive-energy (negative-energy) projections
only, while $W^{\mu\nu}_C$ is a crossed term containing products of both positive- and negative-energy projections. Following the
general arguments presented in \cite{Caballero98a} and introducing the functions associated with the upper/lower components
in the bound nucleon wave function in momentum space:
\begin{eqnarray}
 \alpha_k(p) &=& g_k(p)-\frac{p}{\overline{E}+M_N}S_k f_k(p)\, ,\label{alpha_k}\\
 \beta_k(p) &=& \frac{p}{\overline{E}+M_N}g_k(p)-S_k f_k(p)\, ,\label{beta_k}
\end{eqnarray}
we can finally express the different contributions to the hadronic tensors in the form:
\begin{eqnarray}
W^{\mu\nu}_{P} &=& e^2
 \underbrace{\frac{\overline{E}+M_N}{M_N}
 \frac{|\alpha_k(p)|^2}{16\pi}}_{N_{uu}(p)}
 \underbrace{\sum_{s_Ns}\biggl[\overline{u}({\bf p}_N,s_N)
 \hat{J}^{\mu}_{EM}u({\bf p},S)\biggr]^*
 \biggl[\overline{u}
 ({\bf p}_N,s_N)\hat{J}^{\nu}_{EM}
 u({\bf p},S)\biggr]}_{{\cal W}^{\mu\nu}}\nonumber\\
&=& e^2N_{uu}(p)\ {\cal W}^{\mu\nu}\,, \label{WmunuP}
\end{eqnarray}
\begin{eqnarray}
W^{\mu\nu}_{N} &=& 
e^2\underbrace{\frac{\overline{E}+M_N}{M_N}\frac{|\beta_k(p)|^2}{16\pi}}_{N_{vv}(p)} 
\underbrace{\sum_{s_Ns}
\biggl[\overline{u}({\bf p}_N,s_N)\hat{J}^{\mu}_{EM}v({\bf p},S)\biggr]^*
\biggl[\overline{u}({\bf p}_N,s_N)\hat{J}^{\nu}_{EM}
v({\bf p},S)\biggr]}_{{\cal Z}^{\mu\nu}}\nonumber\\
&=& e^2N_{vv}(p)\ {\cal Z}^{\mu\nu}\,, \label{WmunuN}
\end{eqnarray}
\begin{eqnarray}
 W^{\mu\nu}_{C} &=& 
 e^2 \overbrace{\left(-\frac{\overline{E}+M_N}{M_N}\right)
 \frac{\alpha_k(p)\beta_k(p)}{8\pi}}^{N_{uv}(p)}\sum_{ss'}\frac{1}{2}
 \left\lbrace \langle s'|\frac{{\bf \sigma}\cdot{\bf p}}{p}|s\rangle
 \right.
 \nonumber\\
%%%%%% 
&\times& \sum_{s_N}\biggl[ 
\bigl[\overline{u}({\bf p}_N,s_N)\hat{J}^{\mu}_{EM}u({\bf p},S)\bigr]^* 
\bigl[\overline{u}({\bf p}_N,s_N)\hat{J}^{\nu}_{EM}v({\bf p},S')\bigr]
\biggr.\nonumber\\
%%%%%%
&+& \left.\biggl. 
\bigl[\overline{u}({\bf p}_N,s_N)\hat{J}^{\mu}_{EM}v({\bf p},S)\bigr]^* 
\bigl[\overline{u}({\bf p}_N,s_N)\hat{J}^{\nu}_{EM}u({\bf p},S')\bigr] 
\biggr]\right\rbrace\nonumber\\
&=& e^2 N_{uv}(p)\ {\cal N}^{\mu\nu}\,. \label{WmunuC}
\end{eqnarray}

A similar decomposition holds for the WNC hadronic tensor:
\begin{eqnarray}
\widetilde{W}^{\mu\nu}_{P} &=& \left(\frac{eg}{4\cos\theta_W}\right) 
 N_{uu}(p)\ \widetilde{{\cal W}}^{\mu\nu}\\
\widetilde{W}^{\mu\nu}_{N} &=& \left(\frac{eg}{4\cos\theta_W}\right) 
N_{vv}(p)\ \widetilde{{\cal Z}}^{\mu\nu}\\
\widetilde{W}^{\mu\nu}_{C} &=& \left(\frac{eg}{4\cos\theta_W}\right) 
 N_{uv}(p)\ \widetilde{{\cal N}}^{\mu\nu} \, ,
\end{eqnarray}
where the single-nucleon tensors $\widetilde{{\cal W}}^{\mu\nu}$, $\widetilde{{\cal Z}}^{\mu\nu}$ and $\widetilde{{\cal N}}^{\mu\nu}$ 
are defined in a similar way to the purely EM ones (${\cal W}^{\mu\nu}$, ${\cal Z}^{\mu\nu}$, ${\cal N}^{\mu\nu}$) given 
in Eqs.~(\ref{WmunuP}), (\ref{WmunuN}), (\ref{WmunuC}), but inter-changing one of the purely EM nucleon current matrix element with the corresponding WNC one.\\

%%%%%%%%%%%%%%%%%%%%%%%%%%%%%%%%%%%%%%%%%%%%%%%%%%%%%%%%%%%
%%%%%%%%%%%%%%%%%%%%%%%%%%%%%%%%%%%%%%%%%%%%%
%%%%%%%%%%%%%%%%%%%%%%%%%%%%%%%%%%%%%%%%%%%%%%%%%%%%%%%%%%%
%%%%%%%%%%%%%%%%%%%%%%%%%%%%%%%%%%%%%%%%%%%%%

In what follows we evaluate the explicit expressions for the {\bf PV single-nucleon tensors}. We consider both CC1 and CC2 prescriptions for the
vector part in the weak current. The purely EM tensors have been already presented in previous work~\cite{Caballero98a}, hence here we restrict our
attention to the WNC tensors. To simplify the analysis we separate the positive, negative and crossed contributions.

\subsection{Positive-energy tensor: $uu$ contribution}

In this case the interference single-nucleon tensor is given by the following trace:
\begin{eqnarray}
 \widetilde{{\cal W}}^{\mu\nu} &=&  
 \bigl[\widetilde{{\cal W}}^{\mu\nu}_{V} + \widetilde{{\cal W}}^{\mu\nu}_{A} \bigr]\nonumber\\ 
 &=& \frac{1}{4M_N^2} 
 \text{Tr}\bigl[ 
 \overline{\hat{J}_{EM}^{\mu}} 
 (\pnslash+M_N)\left(\hat{J}^{\nu}_{WNC,V} + \hat{J}^{\nu}_{WNC,A}\right)
 (\pbslash+M_N)\bigr]\, .
\end{eqnarray}
The following expressions are obtained for the two prescriptions of the vector term in the weak current:
\begin{itemize}
%%%%%%
\item {\bf CC1 Vector contribution:}
\begin{eqnarray}
 \widetilde{{\cal W}}^{\mu\nu}_{V} &=& 
 \widetilde{{\cal S}}^{\mu\nu}_{V,uu}=\frac{1}{M_N^2}
 \Biggl\lbrace (F_1+F_2)(\widetilde{F}_1+\widetilde{F}_2) 
 \left(\overline{P}^{\mu}P^{\nu}_N+\overline{P}^{\nu}P_N^{\mu}+
   \frac{\overline{Q}^2}{2}g^{\mu\nu}\right) \Biggr. \nonumber\\
   &+&\frac{1}{2} \bigg[ F_2\widetilde{F}_2\left(1-\frac{\overline{Q}^2}{4M_N^2}\right) 
   -F_2(\widetilde{F}_1+\widetilde{F}_2) - \widetilde{F}_2(F_1+F_2) \biggr]
   C^{\mu}C^{\nu}\Biggr\rbrace\,.
\end{eqnarray}
%%%%%%
%%%%%%
%%%%%%
\item {\bf CC2 Vector contribution:}
Contrary to the previous case, here the tensor has both symmetric and antisymmetric parts:
\begin{eqnarray}
    \widetilde{{\cal W}}^{\mu\nu}_{V} = \widetilde{{\cal S}}^{\mu\nu}_{V,uu} 
				      + \widetilde{{\cal A}}^{\mu\nu}_{V,uu}\,.
\end{eqnarray}
The symmetric tensor is
\begin{eqnarray}
\widetilde{{\cal S}}^{\mu\nu}_{V,uu}&=&\frac{1}{M_N^2} 
  \Biggl\lbrace F_1\widetilde{F}_1\left[\overline{P}^{\mu}P_N^{\nu}
  +\overline{P}^{\nu}P_N^{\mu}+
  \frac{\overline{Q}^2}{2} g^{\mu\nu}\right] \Biggr.
%%%%%%%%%  
+\frac{F_1\widetilde{F}_2 + F_2\widetilde{F}_1}{2} 
     g^{\mu\nu}\overline{Q}\cdot Q\nonumber\\
%%%%%%%%%%%%%%%%%
&-&\left(\frac{F_1\widetilde{F}_2}{4}
\left(\overline{Q}^{\mu}Q^{\nu}+\overline{Q}^{\nu}Q^{\mu}\right)
+\frac{F_2\widetilde{F}_1}{4}
\left(\overline{Q}^{\nu}Q^{\mu}+\overline{Q}^{\mu}Q^{\nu}\right)\right)
\nonumber\\
%%%%%%%%%%%%%%%%%%
&+& \frac{F_2 \widetilde{F}_2}{4M_N^2} 
\biggl[ P_N\cdot Q(\overline{P}^{\nu}Q^{\mu}+\overline{P}^{\mu}Q^{\nu})+\overline{P} 
\cdot Q (P_N^{\mu}Q^{\nu}+P_N^{\nu}Q^{\mu}) \biggr.\nonumber\\
&-& Q^2(P_N^{\mu}\overline{P}^{\nu}+P_N^{\nu}\overline{P}^{\mu}) - 
Q^{\mu}Q^{\nu} \left(2M_N^2-\frac{\overline{Q}^2}{2}\right)\nonumber\\
%%%%%%%%%%%%%%%%
&+&  \left. g^{\mu\nu} 
\left( Q^2\left(2M_N^2-\frac{\overline{Q}^2}{2}\right)-2(P_N\cdot Q)
(\overline{P}\cdot Q)\right)\right]\Biggr\rbrace\,,
\end{eqnarray}
and the antisymmetric one is
\begin{eqnarray}
\widetilde{{\cal A}}^{\mu\nu}_{V,uu}=-\left(\frac{F_1\widetilde{F}_2}{4}
\left(\overline{Q}^{\mu}Q^{\nu}-\overline{Q}^{\nu}Q^{\mu}\right)
+\frac{F_2\widetilde{F}_1}{4}
\left(\overline{Q}^{\nu}Q^{\mu}-\overline{Q}^{\mu}Q^{\nu}\right)\right)\,.
\end{eqnarray}
%%%%%%
%%%%%%
%%%%%%
\item {\bf CC1 Axial contribution:}
\begin{eqnarray}
  \widetilde{{\cal W}}^{\mu\nu}_{A} = 
  \widetilde{{\cal A}}^{\mu\nu}_{A,uu}=\frac{i}{M_N^2}(F_1+F_2)G_A^e\ 
 \epsilon^{\mu\nu\alpha\beta}\overline{P}_{\alpha}P_{N,{\beta}}  \, . 
\label{cc1axialP}
\end{eqnarray}
%%%%%%
%%%%%%
%%%%%%
\item {\bf CC2 Axial contribution:}
Again symmetric and antisymmetric parts contribute to the whole single-nucleon tensor:
\ba
\widetilde{{\cal W}}^{\mu\nu}_{A} = \widetilde{{\cal S}}^{\mu\nu}_{A,uu}
				  +\widetilde{{\cal A}}^{\mu\nu}_{A,uu} \, 
\ea
with
\begin{eqnarray}
 \widetilde{{\cal S}}^{\mu\nu}_{A,uu} &=& \frac{i}{M_N^2}\Biggl\lbrace
 \frac{F_2}{4M_N^2}\widetilde{G}_p
 \left(Q^\nu\epsilon^{\mu\alpha\beta\delta} + Q^\mu\epsilon^{\nu\alpha\beta\delta}\right)
 \overline{P}_\alpha P_{N,\beta} Q_\delta\Biggr\rbrace\,.
\end{eqnarray}
\begin{eqnarray}
 \widetilde{{\cal A}}^{\mu\nu}_{A,uu} &=& \frac{i}{M_N^2}\Biggl\lbrace 
 \epsilon^{\mu\nu\alpha\beta} 
 G_A^e \left[\frac{F_2}{2}(P_N+\overline{P})_{\alpha}Q_{\beta} 
 + F_1\overline{P}_{\alpha}P_{N,\beta}\right] \phantom{\Biggr\rbrace}\nonumber\\
 \phantom{\Biggl\lbrace} 
 &+&\frac{F_2}{4M_N^2}\widetilde{G}_p
 \left(Q^\nu\epsilon^{\mu\alpha\beta\delta} - Q^\mu\epsilon^{\nu\alpha\beta\delta}\right)
 \overline{P}_\alpha P_{N,\beta} Q_\delta \Biggr\rbrace\,.
\end{eqnarray}
\end{itemize}

%%%%%%%%%%%%%%%%%%%%%%%%%%%%%%%%%%%%%%%%%%%%%%%%%%%%%%%%%%%%%%%%%%%%%%
%%%%%%%%%%%%%%%%%%%%%%%%%%%%%%%%%%

\subsection{Negative-energy tensor: $vv$ contribution}

In this case the general expression for the single-nucleon tensor in terms of traces reads:
\begin{eqnarray}
 \widetilde{{\cal Z}}^{\mu\nu} =  
 \frac{1}{4M_N^2} 
 \text{Tr}
 \bigl[ \overline{\hat{J}_{EM}^{\mu}}
 (\pnslash+M_N)\hat{J}^{\nu}_{WNC}(\pbslash - M_N)\bigr]\, .
\end{eqnarray}
The following explicit expressions are obtained:
\begin{itemize}
 %%%%%%%
 \item {\bf CC1 Vector contribution}
It presents symmetric and antisymmetric terms, 
\begin{eqnarray}
\widetilde{{\cal Z}}^{\mu\nu}_{V}=\widetilde{{\cal S}}^{\mu\nu}_{V,vv}
				 + \widetilde{{\cal A}}^{\mu\nu}_{V,vv}\, 
\end{eqnarray}
with
\begin{eqnarray}
\widetilde{{\cal S}}^{\mu\nu}_{V,vv} &=& \frac{1}{M_N^2}
 \Biggl\lbrace  (F_1+F_2)(\widetilde{F}_1+\widetilde{F}_2)
 \left(\overline{P}^{\mu}P^{\nu}_N+\overline{P}^{\nu}P_N^{\mu} -
   \frac{(\overline{P}+P_N)^2}{2}g^{\mu\nu}\right)\nonumber\\
&-&  \frac{F_2 \widetilde{F}_2}{8M_N^2}\overline{Q}^2
(\overline{P}+P_N)^{\mu}(\overline{P}+P_N)^{\nu}\nonumber\\
 &+&\frac{\widetilde{F}_2(F_1+F_2)+F_2(\widetilde{F}_1+\widetilde{F}_2)}{2}
 (P_N^{\mu}P_N^{\nu}-\overline{P}^{\mu}\overline{P}^{\nu})\Biggr\rbrace \,,
\end{eqnarray}
\begin{eqnarray}
\widetilde{{\cal A}}^{\mu\nu}_{V,vv} &=& \frac{1}{M_N^2}
 \Biggl\lbrace\frac{\widetilde{F}_2F_1-F_2\widetilde{F}_1}{2}
 \left(P_N^\mu\overline{P}^\nu-P_N^\nu\overline{P}^\mu \right) \Biggr\rbrace \,.
\end{eqnarray}
 %%%%%%%
 %%%%%%%
 %%%%%%%
 \item {\bf CC2 Vector contribution}
In this case the symmetric and antisymmetric parts of the tensor are
\begin{eqnarray}
  \widetilde{{\cal S}}^{\mu\nu}_{V,vv} &=&
  \frac{1}{M_N^2} \Biggl\lbrace F_1\widetilde{F}_1 
  \left(\overline{P}^{\mu}P_N^{\nu}+\overline{P}^{\nu}P_N^{\mu} 
  - \frac{(\overline{P}+P_N)^2}{2} g^{\mu\nu}\right) \Biggr.\nonumber\\
%%%%%%%%%%%%%%%%%%%%%%%%%%%%%%%%%%%%%%%%
 %%%%%%%%%%%%%%%%%%%%%%%%%%%
&+& \frac{F_1 \widetilde{F}_2 + F_2 \widetilde{F}_1}{4}
\left(Q^{\mu}(\overline{P}+P_N)^{\nu}+Q^{\nu}(\overline{P}+P_N)^{\mu}
- 2Q\cdot(\overline{P}+P_N)g^{\mu\nu}\right)
\nonumber\\
%%%%%%%%%%%%%%%%%%%%%%%%%%%%%%%%%%%%%%%%
&+& \frac{F_2\widetilde{F}_2}{4M_N^2} 
\biggl[ P_N\cdot Q(\overline{P}^{\mu}Q^{\nu}+\overline{P}^{\nu}Q^{\mu})+\overline{P} 
\cdot Q (P_N^{\mu}Q^{\nu}+P_N^{\nu}Q^{\mu})
+ \frac{\overline{Q}^2}{2}Q^{\mu}Q^{\nu}\biggr.\nonumber\\
%%%%%%%%%%%%%%%%%%%%%%%%%%%%%%%%%%%%%%%%
&-& \left. Q^2 (P_N^{\mu}\overline{P}^{\nu}+P_N^{\nu}\overline{P}^{\mu}) - g^{\mu\nu} \left(\frac{\overline{Q}^2Q^2}{2} 
+ 2(P_N\cdot Q)(\overline{P}\cdot Q)\right)\right]\Biggr\rbrace\,,
\end{eqnarray}
\begin{eqnarray}
  \widetilde{{\cal A}}^{\mu\nu}_{V,vv}=
  \frac{1}{M_N^2} \Biggl\lbrace 
\frac{F_1 \widetilde{F}_2 - F_2 \widetilde{F}_1}{4}
\left(Q^{\mu}(\overline{P}+P_N)^{\nu}-Q^{\nu}(\overline{P}+P_N)^{\mu}\right)
\Biggr\rbrace\,.
\end{eqnarray}
 %%%%%%%
 %%%%%%%
 %%%%%%%
  \item {\bf CC1 Axial contribution}
Here the result coincides with the expression already obtained for the CC1 axial contribution in the case of the purely positive-energy tensor in Eq.~(\ref{cc1axialP}). 
 %%%%%%%
 %%%%%%%
 \item {\bf CC2 Axial contribution}
Its symmetric and antisymmetric parts, $\widetilde{{\cal Z}}^{\mu\nu}_{A} = \widetilde{{\cal S}}^{\mu\nu}_{A,vv}+\widetilde{{\cal A}}^{\mu\nu}_{A,vv}$, are
given by
\begin{eqnarray}
  \widetilde{{\cal S}}^{\mu\nu}_{A,vv} &=& \frac{i}{M_N^2}\Biggl\lbrace 
  \frac{F_2\widetilde{G}_p}{4M_N^2}(Q^\nu\epsilon^{\mu\alpha\beta\delta}
 + Q^\mu\epsilon^{\nu\alpha\beta\delta})
 \overline{P}_\alpha P_{N,\beta} Q_\delta\Biggr\rbrace\,,
\end{eqnarray}
\begin{eqnarray}
  \widetilde{{\cal A}}^{\mu\nu}_{A,vv} &=& \frac{i}{M_N^2} 
 \Biggl\lbrace\epsilon^{\mu\nu\alpha\beta} 
 G_A^e \left[\frac{F_2}{2}(\overline{P}-P_N)_{\alpha}Q_{\beta} 
 + F_1 \overline{P}_{\alpha}P_{N,\beta}\right] \phantom{\Biggr\rbrace}\nonumber\\
 \phantom{\Biggl\lbrace} 
 &+&\frac{F_2\widetilde{G}_p}{4M_N^2}(Q^\nu\epsilon^{\mu\alpha\beta\delta}
 -Q^\mu\epsilon^{\nu\alpha\beta\delta})
 \overline{P}_\alpha P_{N,\beta} Q_\delta\Biggr\rbrace\,.
\end{eqnarray}
 %%%%%%%
 %%%%%%%
 %%%%%%%
\end{itemize}

\subsection{Crossed tensor: $uv$ contribution}

The non-diagonal spin single-nucleon tensor ${\cal N}^{\mu\nu}$ that enters in the evaluation of the crossed $uv$ hadronic
tensor can be written in terms of a diagonal tensor constructed from spinors quantized with respect to a spin axis pointing
along a generic direction, ${\cal R}^{\mu\nu}(\theta_R,\phi_R)$. 
This is the spin precession technique that has presented
in detail in~\cite{Caballero93,Caballero98a}. 
Here we simply summarize the main results applied specifically to the case of the
PV response functions. 

In general we can write
\be
 {\cal N}^{\mu\nu}= {\cal R}^{\mu\nu}(0,0)\cos\theta + 
 \left({\cal R}^{\mu\nu}(\frac{\pi}{2},0)\cos\phi + 
 {\cal R}^{\mu\nu}(\frac{\pi}{2},\frac{\pi}{2})\sin\phi_N\right)\sin\theta \, ,
\ee
where $\theta$, $\phi_N$ are the angles defining the direction of the bound nucleon momentum ${\bf p}$ and the tensor 
${\cal R}^{\mu\nu}$ is given in the general form
\be
 {\cal R}^{\mu\nu}(\theta_R,\phi_R)=\frac{1}{4M_N}\text{Tr}
 \bigl[\slslash\overline{\hat{J}^{\mu}_{EM}}(\pnslash + M_N)
 \hat{J}^{\nu}_{WNC}\bigr]\, 
\ee
with $\hat{J}^\mu_{EM}$ ($\hat{J}^\mu_{WNC}$) the purely EM (WNC) current operators. Isolating the vector and axial contributions in 
the WNC operator, and after some algebra, the following results are obtained for the two prescriptions of the vector current:
\subsubsection{Vector interference contribution}
%%%%%%%
\begin{itemize}
 \item {\bf CC1 prescription:}
The single-nucleon tensor has symmetric and antisymmetric parts,
\be
 \widetilde{{\cal R}}_V^{\mu\nu} = \widetilde{{\cal S}}_{V,uv}^{\mu\nu}
				 + \widetilde{{\cal A}}_{V,uv}^{\mu\nu}\,
\ee
with
\begin{eqnarray}
 \widetilde{{\cal S}}_{V,uv}^{\mu\nu} &=& \frac{1}{M_N}
 \Biggl\lbrace (F_1+F_2)(\widetilde{F}_1+\widetilde{F}_2)
 \biggl(S_L^{\mu}P_N^{\nu}+S_L^{\nu}P_N^{\mu}-(P_N\cdot S_L)g^{\mu\nu}\biggr)
 \Biggr.\nonumber\\
 %%%%%%%%%%%%%%%%%%%%%%%%%%%%%%%%%%
&+& \frac{F_2\widetilde{F}_2}{4M_N^2}(P_N\cdot S_L)(\overline{P}+P_N)^{\mu}
(\overline{P}+P_N)^{\nu}\nonumber\\
%%%%%%%%%%%%%%%%%%%%%%%%%%%%%%%%%%%
&-& \frac{F_2(\widetilde{F}_1+\widetilde{F}_2) + \widetilde{F}_2(F_1+F_2) }{4}
\left((\overline{P}+P_N)^{\mu}S_L^{\nu}+(\overline{P}+P_N)^{\nu}S_L^{\mu}\right)
\Biggr\rbrace \,, \nonumber \\
&&
\end{eqnarray}
and
\begin{eqnarray}
 \widetilde{{\cal A}}_{V,uv}^{\mu\nu} &=& \frac{1}{M_N}\Biggl\lbrace 
 -\frac{F_2(\widetilde{F}_1+\widetilde{F}_2) - \widetilde{F}_2(F_1+F_2) }{4}
\left((\overline{P}+P_N)^{\mu}S_L^{\nu}-(\overline{P}+P_N)^{\nu}S_L^{\mu}\right)
\Biggr\rbrace \, .\nonumber\\ 
\end{eqnarray}
%%%%%%%%
%%%%%%%%
%%%%%%%%
\item {\bf CC2 prescription:}
Likewise, the symmetric and antisymmetric parts result:
\begin{eqnarray}
 \widetilde{{\cal S}}_{V,uv}^{\mu\nu} &=& \frac{1}{M_N}
 \Biggl\lbrace F_1\widetilde{F}_1\biggl(S_L^{\mu}P_N^{\nu}+S_L^{\nu}P_N^{\mu}
 -(P_N\cdot S_L)g^{\mu\nu}\biggr)\Biggr.\nonumber\\
 %%%%%%%%%%%%%%%%%%%%%%%%%%
&-& \frac{F_1\widetilde{F}_2 +\widetilde{F}_1F_2}{2}(Q\cdot S_L) g^{\mu\nu}
+ \frac{F_1\widetilde{F}_2+F_2\widetilde{F}_1}{4}
\left(Q^\mu S_L^\nu+Q^\nu S_L^\mu\right)\nonumber\\
%%%%%%%%%%%%%%%%%%%%%%%%%%%
&+& \frac{F_2\widetilde{F}_2}{4M_N^2}\biggl[P_N\cdot Q(S_L^{\mu}Q^{\nu}+S_L^{\nu}Q^{\mu}) 
+ S_L\cdot Q(P_N^{\mu}Q^{\nu}+P_N^{\nu}Q^{\mu})\nonumber\\
%%%%%%%%%%%%%%%%%%%%%%%%%%%
&-& Q^2(P_N^{\mu}S_L^{\nu}+P_N^{\nu}S_L^{\mu})-(P_N\cdot S_L)Q^{\mu}Q^{\nu}\nonumber\\
%%%%%%%%%%%%%%%%%%%%%%%%%%%
&+& g^{\mu\nu}\bigl(Q^2P_N\cdot S_L -2(P_N\cdot Q)
(S_L\cdot Q)\bigr)\biggr]\Biggr\rbrace \,,
\end{eqnarray}
and
\begin{eqnarray}
 \widetilde{{\cal A}}_{V,uv}^{\mu\nu} = \frac{1}{M_N}\Biggl\lbrace
  \frac{F_1\widetilde{F}_2-F_2\widetilde{F}_1}{4}
\left(Q^\mu S_L^\nu-Q^\nu S_L^\mu\right)
\Biggr\rbrace \,.
\end{eqnarray}
\end{itemize}
\subsubsection{Axial interference}
In this case the expressions for the tensor can be cast as follows:
%%%%%%%%
\begin{itemize}
%%%%%%%%
%%%%%%%%
%%%%%%%%
\item {\bf CC1 prescription}
\be
\widetilde{{\cal R}}_A^{\mu\nu} = \widetilde{{\cal A}}_{A,uv}^{\mu\nu}
= -\frac{i}{M_N}(F_1+F_2)G_A^e 
\epsilon^{\mu\nu\alpha\beta}P_{N,\alpha}S_{L,\beta}
\ee
%%%%%%%%
%%%%%%%%
%%%%%%%%
\item {\bf CC2 prescription:}
The symmetric term is
\begin{eqnarray}
\widetilde{{\cal A}}_{A,uv}^{\mu\nu} &=& -\frac{i}{M_N}
\Biggl\lbrace G_A^e\epsilon^{\mu\nu\alpha\beta}
\left(F_1 P_{N,\alpha}
+ \frac{F_2}{2} Q_\alpha \right)S_{L,\beta}\biggr.\nonumber\\
\biggl. &-& \frac{F_2\widetilde{G}_P}{4M_N^2} 
\left(Q^\nu\epsilon^{\mu\alpha\beta\delta}+Q^\mu\epsilon^{\nu\alpha\beta\delta}\right)
Q_\alpha P_{N,\beta} S_{L,\delta}\Biggr\rbrace\,,
\end{eqnarray}
and the antisymmetric one
\begin{eqnarray}
\widetilde{{\cal S}}_{A,uv}^{\mu\nu} = -\frac{i}{M_N}\Biggl\lbrace 
- \frac{F_2\widetilde{G}_P}{4M_N^2} 
\left(Q^\nu\epsilon^{\mu\alpha\beta\delta}-Q^\mu\epsilon^{\nu\alpha\beta\delta}\right)
Q_\alpha P_{N,\beta} S_{L,\delta}\Biggr\rbrace\,.
\end{eqnarray}
%%%%%%%%%%
%%%%%%%%%%
%%%%%%%%%%
\end{itemize}

%%%%%%%%%%%%%%%%%%%%%%%%%%%%%%

% \linespread{0.5}
\appendix
\small
\bibliographystyle{apsrev4-1}

\bibliography{bibliography}

\end{document}